\documentclass[twocolumn,aps,prd,superscriptaddress,showpacs,floatfix,longbibliography,10pt]{revtex4-1}

\usepackage{amsmath}
\usepackage{bm}
\usepackage{graphicx}
\usepackage{url}
\usepackage{cancel}
\usepackage[colorlinks,linkcolor=blue,citecolor=blue,filecolor=black,urlcolor=blue]{hyperref}
\usepackage{booktabs} % 优化表格线条
\usepackage{tabularx} % 自适应列宽
\usepackage{dcolumn}% Align table columns on decimal point
\usepackage[usenames]{color}
\usepackage{amssymb}
\usepackage{slashed}
\usepackage{multirow}
\usepackage{float}
\usepackage{harpoon}
\usepackage{MnSymbol}
\usepackage{appendix}
\usepackage{xcolor}%高亮
\usepackage{cleveref}
\usepackage{ulem} % delete line \sout
\usepackage{subcaption} % 子图标题
\usepackage[justification=raggedright]{caption} % 左对齐

\makeatletter

\newcommand{\ket}[1]{\vert #1\rangle}
\newcommand{\braket}[2]{\langle #1\vert #2\rangle}

\makeatother

\begin{document}

%\title{The $\theta$-term effects on isospin asymmetric hot and dense quark matter}

\title{%
  Chromomagnetic condensation and perturbative confinement induced by imaginary rotation in SU(2) Yang-Mills Theory}
  
  %Imaginary Rotation and Chromomagnetic Condensation in SU(2) Yang-Mills Theory
%}
\author{Lei Zhang}\email[Correspond to\ ]{zhanglei231@mails.ucas.ac.cn}
\affiliation{School of Physical Sciences, University of Chinese Academy of Sciences, Beijing 101408, China}
\author{Kun Xu}\email[Correspond to ]{xukun@bit.edu.cn}
\affiliation{School of Physics, Beijing Institute of Technology, Beijing, 100049,
  P.R. China}
\author{Mei Huang}\email[Correspond to\ ]{huangmei@ucas.ac.cn}
\affiliation{School of Nuclear Science and Technology, University of Chinese Academy of Sciences, Beijing, 101408, P.R. China}
\date{\today}

\begin{abstract}
We perturbatively investigate the rotation effect on the Polyakov loop potential in SU(2) gauge theroy within a chromomagnetic background. It is observed that the imaginary rotation spontaneously induces both confinement and chromomagnetic condensation at high temperatures, thereby provides a perturbative window to explore non-perturbative dynamics. Compared to the case without including the induced chromomagnetic field, the perturbative confinement transition becomes first-order, with a temperature-dependent phase boundary that asymptotically approaches $\tilde{\Omega}_c = \pi/\sqrt{3}$ at high temperatures. This leads to a significantly enriched $\tilde{\Omega}$-$T$ phase diagram characterized by an expanded deconfined region. For real angular velocities, we find that the chromomagnetic condensate decreases with increasing rotation, and that the coupling between rotation, spin, and the chromomagnetic background leads to a cusp in the Polyakov loop potential, suggesting that the underlying dynamics could be more intricate.
\end{abstract}

\maketitle
\section{
Introduction}
\label{sec:intro}
In $\mathrm{SU}(N)$ theories, the negative sign of the $\beta$ function in the renormalization group flow, manifesting asymptotic freedom at ultraviolet(UV). In the infrared, the coupling grows without bound within perturbation theory, signaling the onset of a strongly coupled confining phase where perturbative methods fail. 
The non-perturbative effects in $\mathrm{SU}(N)$ theories—confinement and chiral symmetry breaking—have long been central themes in the study of strong interactions. The most significant non-perturbative property is the existence of a deconfinement phase transition at finite temperature, investigated through non-perturbative methods such as semi-classical contributions \cite{Davies:1999uw,Poppitz:2012sw,Aitken:2017ayq,Polyakov:1976fu}, lattice regularization \cite{Wilson:1974sk,Polyakov:1978vu,Boyd:1996bx}, functional renormalization group \cite{Braun-Munzinger:2008szb}, Dyson-Schwinger equations \cite{Alkofer:2007zb}, and phenomenological effective models \cite{Fukushima:2003fw}. These approaches have yielded many crucial insights.

In the heavy quark limit, the order parameter describing the deconfinement phase transition is the
Polyakov loop in the fundamental representation, which is the trace of a Wilson line winding around the compactified Euclidean time direction. The modulus (or magnitude) of the Polyakov loop $|\langle L \rangle|$ for simplicity denoted as $| L |$ represents the effect of a static color source at finite temperature: its expectation value gives the free energy required to insert an infinitely heavy, static test charge in the fundamental representation into a thermal medium of gluons, i.e.,
\begin{equation}
    | L | \propto e^{-\beta F_q}, \qquad | L | \in [0, 1].
\end{equation}
$\mathrm{SU}(N)$ gauge theories possess a $\mathbb{Z}(N)$ center symmetry. This symmetry is intimately connected to the behavior of the Polyakov loop under such transformations, making it the order parameter for this symmetry. Thus, the Polyakov loop tightly links the physical phenomenon of confinement to the mathematical framework of center symmetry and its spontaneous breaking. In the confined phase: the free energy required to pull an isolated static quark from the vacuum is infinite, corresponding to $| L |= 0$, and the $\mathbb{Z}(N)$ symmetry is unbroken. In the deconfined phase: color charge can be screened by the thermal medium, allowing a single static quark to exist with finite free energy, corresponding to $| L | \neq 0$, and the $\mathbb{Z}(N)$ symmetry is spontaneously broken. The magnitude of $| L |$ directly reflects the strength of the screening.

As a key component of the QCD phase structure, the behavior of the deconfinement phase transition is indeed regulated by various external parameters. Examples include the inverse magnetic catalysis effect under external magnetic fields \cite{Bali:2011qj,Endrodi:2015oba,Gatto:2010pt} and the search for the Critical End Point (CEP) at finite density \cite{Fischer:2011mz,Fu:2019hdw,Fukushima:2008wg}. In non-central relativistic heavy-ion collisions, a nuclear system is created with significant rotation and high temperature \cite{STAR:2017ckg,Fukushima:2018grm,Becattini:2020ngo,Huang:2020dtn}, where the vorticity $\omega$ of the created matter can be as large as $10^{22} \text{s}^{-1}$. The effect of rotation $\omega$ has attracted widespread theoretical and experimental attention, becoming an important external parameter in quantum field theory research, alongside temperature, chemical potential, and magnetic field.

Rotation effects have been extensively studied in various systems including complex scalar fields \cite{Bai:2025rwx,Guo:2021gbz,Siri:2024scq,Salvio:2025rma}, fermion fields \cite{Ayala:2021osy,Chen:2015hfc,Jiang:2016wvv,Gaspar:2023nqk,Salvio:2025ggj}, and vector fields \cite{Xu:2021kjg,Wei:2023pdf,Salvio:2026ewl}. Such effects exist in diverse systems; for instance, heavy nuclei with spontaneously broken rotational symmetry can restore it under external rotation, generating rotational bands \cite{Bohr:1976zz,Otsuka:2019diq}. In QCD, rotation leads to important phenomena such as the chiral vortical effect \cite{Vilenkin:1978hb,Vilenkin:1979ui,Vilenkin:1980zv,Son:2009tf}. A fundamental inquiry then arises: how the QCD phase structure evolves under the influence of $\omega$, particularly regarding the chiral phase transition \cite{Chen:2015hfc,Jiang:2016wvv,Ebihara:2016fwa,Chernodub:2016kxh,Chernodub:2017ref,Wang:2018sur,Chen:2022mhf}. Recently, rotating gluonic systems and the deconfinement phase transition have garnered significant attention, stemming from the larger spin polarization of gluons and the dominance of soft gluons in the Quark-Gluon Plasma (QGP). The deconfinement phase transition in QCD under rotation has also drawn focus \cite{Fujimoto:2021xix,Braguta:2020biu,Braguta:2021jgn,Chen:2020ath,Chernodub:2020qah,Jiang:2023zzu,Jiang:2024zsw,Sun:2024anu,Chen:2024jet,Wang:2025mmv,Fukushima:2025hmh}. However, in rotating systems, challenges arise for lattice calculations from both the consistency between rotational effects and finite density effects \cite{Chen:2015hfc,Jiang:2016wvv} and the fact that rotation makes the pure gauge action complex, introducing a new sign problem. Alternatively, performing lattice QCD simulations by introducing an imaginary angular velocity ($\Omega=-i\omega$) is feasible and has yielded many interesting results \cite{Chernodub:2020qah,Braguta:2020biu,Braguta:2021jgn}. However, most model results contradict these findings.

As outlined above, research on the deconfinement phase transition is fundamentally based on non-perturbative methods. Perturbative calculations of the effective potential for $L$ show that the $\mathbb{Z}(N)$ symmetry is broken at high temperatures, but cannot provide information about its restoration at low temperatures and its relation to confinement \cite{Gross:1980br,Wilson:1974sk,Weiss:1980rj}. However, a recent calculation demonstrates that pure Yang-Mills (YM) theory at sufficiently large $\Omega$ undergoes a confinement phase transition even perturbatively at the rotation center $r=0$. This implies that imaginary rotation $\Omega$ provides a unique framework to manifest non-perturbative physical results within the perturbative regime, which is quite intriguing in itself \cite{Chen:2022smf}.

Notably, in previous studies, a constant chromomagnetic field – the Savvidy vacuum – as the simplest form of non-perturbative vacuum, allows for the analytical evaluation of the one-loop functional determinant \cite{Matinyan:1976mp,Savvidy:1977as,Nielsen:1978rm,Meisinger:1997jt,Haber:1981tr,Haber:1981ts,Actor:1986zf,Actor:1987cf,Meisinger:2002ji}. Its perturbative prediction shows a non-trivial minimum at $H \neq 0$, indicating the existence of chromomagnetic condensation in the vacuum state of non-Abelian gauge theories. While lattice calculations have not found any long-range order for chromomagnetic condensation, the non-zero gluon condensate $\langle G^a_{\mu\nu}G^{a\mu\nu}\rangle$ implies its existence locally \cite{Boyd:1996bx}. More crucially, the coupling of the Polyakov loop to a gauge field condensate exhibits complex interplay. Locally at the rotation center $r=0$, a directed constant color-magnetic background, together with the spin-rotation coupling, modulates the single-particle spectrum, while the chromoelectric background $\phi$ acts as an imaginary chemical potential affecting statistical distribution. The interplay between these three factors may give rise to new minima, cusps, or imaginary components in the effective potential, ultimately leading to a more complex phase structure.

It is worth noting that Ref.~\cite{Fukushima:2020ncb} pointed out that, for a fermionic system coupled to a constant external magnetic field under rotation, there exists a global thermodynamic-limit problem. However, if the true local chromomagnetic background is understood as a slowly varying field configuration within a small region, then a globally uniform background is precisely the lowest-order approximation to such a local slowly varying background. Therefore, the more appropriate interpretation of our calculation is that it is an analysis for a local, slowly varying chromomagnetic background, rather than a strict bulk thermodynamic construction for a globally uniform chromomagnetic medium in the whole infinite space.

In this work, we derive the one-loop effective potential for pure $\mathrm{SU}(2)$ theory under rotation in the presence of chromoelectric and chromomagnetic background fields. We identify two components within the effective potential: one promotes chromomagnetic condensation and another suppresses it. In the perturbative regime, our results show strong suppression of chromomagnetic condensation. However, the inclusion of imaginary rotation introduces a novel competitive relationship between these two potential components, leading to an explicit breaking of $\mathbb{Z}(2)$ symmetry. This causes significant changes in the region of perturbative confinement, the order of the phase transition, and the temperature dependence of the phase boundary. The adiabatic continuity from the perturbative confinement to the confined phase at low temperature might be affected along its path in the phase diagram, also suggesting that at larger imaginary rotations, the critical temperature for deconfinement in the non-perturbative regime may exhibit a non-monotonic dependence on the imaginary rotation. We also address subtleties in the analytical continuation from imaginary to real angular velocity by imposing a causality bound.

This paper is organized as follows. In Sec.~\ref{sec:theory}, we develop the formalism required for evaluating the one-loop effective potential. In Sec.~\ref{sec:3}, we calculate the Polyakov loop potential at different temperatures, presenting the complex coupling phenomenon between the Polyakov loop and chromomagnetic condensation. We analyze the subtle competitive relationship between the chromomagnetic-favoring potential $V_H$ and the chromomagnetic-suppressing potential $V_{\text{nonH}}$ under imaginary rotation, predicting the order of the rotation-induced perturbative confinement phase transition and the temperature dependence of the phase boundary. In Sec.~\ref{sec:4}, we compute the impact of different imaginary rotations on the Polyakov loop potential and obtain the phase diagram for the perturbative confinement phase. In Sec.~\ref{sec:5}, we investigate the effects of real rotation on chromomagnetic condensation and the Polyakov loop potential by imposing boundary conditions. We conclude and provide an outlook in Sec.~\ref{sec:6}.

\section{Polyakov loop potential with imaginary rotation}
\label{sec:theory}
In this section, we introduce the formalism necessary to evaluate the one-loop effective potential for $\mathrm{SU}(2)$ gauge bosons at finite temperature propagating in a special class of background field configurations.

The quadratic fluctuation action in the background field $B_\mu^a=B_\mu\delta^{a3}$ of the $\mathrm{SU}(2)$ Yang-Mills theory is given by:
\begin{equation}
    \mathcal{L}_\text{1-loop}=-\frac{1}{2} \hat{Q}^+\cdot\hat{Q}^{-}-igQ^+_\mu Q^-_\nu B^{\mu\nu}-\frac{1}{4}\tilde{Q}^3\cdot\tilde{Q}^{3}-\frac{1}{4}B\cdot B,
\end{equation}
with
\begin{equation}
    \hat{Q}^+_{\mu\nu}=D_{[\mu}Q^+_{\nu]},\quad \tilde{Q}^3_{\mu\nu}=\partial_{[\mu}Q^3_{\nu]},\quad B_{\mu\nu}=\partial_{[\mu}B_{\nu]}.
\end{equation}
Here, $B$ denotes the background field. It is convenient to identify the fields $Q^\pm=\frac{1}{\sqrt{2}}\left(Q^1\pm i Q^2\right)$ and $Q^3$ as analogous to the $W^\pm$ and $Z^0$ fields, carrying charges $\pm 1$ and  $0$, respectively. The neutral $Z^0$ particle behaves as a free field, while the $W^\pm$ particles behave as charged, spin-1 particles propagating in an external field.

Traditionally, we choose the chromomagnetic field \textit{H} to point along the $z$-direction, which is given by
\begin{equation}
    B_\mu=(B_0,\frac{H}{2}y,-\frac{H}{2}x,0).
\end{equation}
Under the gauge condition $D_{\mu}Q^{\mu,+}=(\partial_{\mu}+igB_{\mu})Q^{\mu,+}=0$, the corresponding gauge-fixing term is:
\begin{equation}
    \mathcal{L}_\text{gf}(B,Q)=-\frac{1}{2}(\partial_\mu Q^{\mu,3})^2-(D_{\mu}Q^{\mu,+})(D_{\mu}^*Q^{\mu,-}).
\end{equation}
the equation of motion (EOM) for $Q_s^+=Q_1^+ +siQ_2^+ $ ($s=\pm$) in the background field is obtained as
\begin{equation}
     (D_0^{2}-\Delta-gH\hat{L}_z+\frac{1}{4}gH^2\rho^2+2sgH )Q_s^+=0.
\end{equation}
The eigenmodes (radial part labeled by $(m,l,s)$) are 
\begin{equation}
    \Phi_{\hat{m}}^{(l)}(\rho,\theta)=\sqrt{\frac{gH}{2\pi}\frac{\hat{m}!}{(|l|+\hat{m})!} }L_{\hat{m}}^{(|l|)}(X)X^{|l|/2}e^{-\frac{1}{2}X}e^{il\theta},
\end{equation}
with the eigenvalues
\begin{equation}
   (E-gB_0)^2=k_z^2+(2m+2s+1)gH.
\end{equation}
where $X=\frac{1}{2}  gH\rho^2$ and $\hat{m}=m+(l-|l|)/2$. Subsequently, we introduce rotation about the $z$-axis with angular velocity $\omega$. According to reference \cite{Xu:2021kjg}, considered vector field  in  the  local  rest/inertial frame, the Lagrangian of the system transforms as follows:
\begin{gather}
    \partial_\mu \rightarrow\partial_\mu+\delta_{\mu 0}R_0,
    \label{rotaM}
\end{gather}
where the operator $R_0$ acts only on the spatial components of a vector field according to 
\begin{equation}
    R_0A_i=-i\omega\hat{L}_zA_i+\omega\epsilon_{i3k}A_k.
\end{equation}
Under this transformation, we find a remarkably simple modification to the EOM of $Q^+_s$:
\begin{equation}
    \partial_0\rightarrow \partial_0-i\omega \left(\hat{L}_z-s\right).
    \label{wn}
\end{equation}
with the gauge-fixing term:
\begin{equation}
    \mathcal{L}_\text{gf}(\partial_0\rightarrow\partial_0-i\omega \hat{L}_z).
\end{equation}
This form reveals a striking symmetry of the chromomagnetic field within this rotating system. In an appropriate gauge, all explicit coupling terms between the rotation $\omega$ and the chromomagnetic field $H$ in the EOM vanish. Consequently, the eigenmodes remain unchanged, while only the eigenvalues become
\begin{equation}
    \left[E-gB_0+(l-s)\omega\right]^2=k_z^2+(2m+2s+1)gH.
\end{equation}
The detailed derivation is provided in Appendix~\ref{A}. It is worth emphasizing that the canonical angular momentum appearing here is not the kinetic angular momentum advocated in Ref.~\cite{Fukushima:2020ncb,Fukushima:2024tkz} to preserve gauge invariance. Instead, because the definition of the gauge transformation for the vector field in the local rest/inertial frame is modified, one can show that the canonical angular momentum appearing here is in fact the crucial ingredient required to maintain gauge invariance. For completeness, we present the detailed proof in Appendix~\ref{D}. It is noteworthy that the form of the chromomagnetic background field remains invariant under rotation. As a vacuum state spontaneously chosen by the physical system, we find that such a configuration of the chromomagnetic field corresponds to a lower-energy state permitted in the rotating system. The Polyakov loop is specified by a constant $B_0$ field, given in the fundamental representation by
\begin{equation}
    B_0\rightarrow-i(g\beta)^{-1}\phi,
\end{equation}
where $\phi \in [0,2\pi]$. The trace of the background Polyakov loop is given by
\begin{equation}
    L = \operatorname{Tr}\mathcal{P} \exp\left( i g \int_0^\beta d\tau \, B_4(\vec{x}, \tau) \right) = 2 \cos(\phi/2).
\end{equation}
Although the effective potential has a tachyonic instability with respect to long-wavelength fluctuations \cite{Nielsen:1978rm}, subsequent studies have shown that, for the unstable modes, including the cubic and quartic terms in the fluctuations yields a real energy density, in agreement with the real part obtained from the quadratic approximation in earlier studies \cite{Kay:2005wm,Parthasarathy:2006is}. Therefore, we can safely obtain the vacuum part and the finite-temperature part of one-loop effective potential:
    \begin{align}
       V_0&=\frac{11(gH)^2}{48\pi^2}\ln\left(\frac{gH}{\Lambda^2}\right), \\
        V_T&=\frac{T}{S}\sum_{m=0}^{\infty}\sum_{s=\pm}\sum_{l=-m}^{N-m}\int_{-\infty}^{+\infty}\frac{dk_z}{2\pi} \notag\\
        &\ln \bigg|1-2\cos{\left[\phi+(l-s)\tilde{\Omega}\right]}e^{-\beta\omega_s}+e^{-2\beta\omega_s}\bigg|.
\label{potentialenergy}
    \end{align}
Here $\tilde{\Omega}=\beta \Omega$, and the background $B_0$ field and the imaginary angular velocity $\Omega$ shift the sum over Matsubara frequencies by a factor of $\phi/\beta+(l-s)\Omega$; $\Lambda$ is a renormalization group-invariant parameter that sets the scale for the gauge theory. In above equation, $S$ is the area of the $xy$-plane, $N$ represents the degeneracy of the Landau level \cite{Chen:2015hfc}, and $\omega_s=\sqrt{k_z^2+(2m+2s+1)gH}$ is the spectrum of the charged $W^\pm$ particles in two polarization states within the field, $m \in \mathbb{N}$ is the Landau quantum number, and $s$ denotes the spin orientation along ($s=+1$) or opposite ($s=-1$) the magnetic field direction (the $s=0$ modes of the gauge field are cancelled by the contribution of the ghost field).

The local thermodynamic potential density $V_T(\vec{r})$ satisfies:
\begin{equation}
    V_T=\frac{1}{V}\int \rho d\rho d\theta dz V_T(\vec{r}),
\end{equation}
yielding:
\begin{align}
     V_T(\vec{r})&=\sum_{s=\pm}\sum_{m=0}^{\infty}u(s,m,\vec{r}),\\
        u(s,m,\vec{r})&=T\sum_{l=-m}^{N-m}|\Phi_{\hat{m}}^{(l)}(\rho,\theta)|^2\int_{-\infty}^{+\infty}\frac{dk_z}{2\pi}\notag\\
        &\ln \bigg|1-2\cos{\left[\phi+(l-s)\tilde{\Omega}\right]}e^{-\beta\omega_s}+e^{-2\beta\omega_s}\bigg|.
\end{align}
In this work, we shall focus on the rotation center $r=0$. Utilizing the property:
\begin{equation}
    \lim_{\vec{r}\rightarrow0 }|\Phi_{n}^{(l)}(\rho,\theta)|^2=\frac{gH}{2\pi}\delta_{l,0},
\end{equation}
we find that at $r=0$, the orbital angular momentum states are frozen, leaving only the spin states. The corresponding expressions are obtained as:
\begin{widetext}
\begin{gather}
    u(-,0,\vec{0})=\frac{gH}{2\beta\pi}\int_{-\infty}^{+\infty}\frac{dk}{2\pi} \ln \Bigg|1-2\cos{(\phi+\tilde{\Omega})}e^{-\beta\sqrt{k^2-gH}}+e^{-2\beta\sqrt{k^2-gH}}\Bigg|,\\
    u(-,1,\vec{0})=\frac{gH}{\beta\pi}\int_{-\infty}^{+\infty}\frac{dk}{2\pi} \operatorname{Re}\ln \left(1-e^{-\beta\sqrt{k^2+gH}+i(\phi+\tilde{\Omega})}\right),\\
    V_{\text{nonH}}=\frac{gH}{\beta\pi}\sum_{m=1}^{\infty}\sum_{s=\pm}\int_{-\infty}^{+\infty}\frac{dk}{2\pi} \operatorname{Re}\ln \left(1-e^{-\beta\omega+i[\phi-s\tilde{\Omega}]}\right),
\end{gather}
\end{widetext}
where $\omega=\sqrt{k^2+(2m+1)gH}$. Here, $V_{\text{nonH}}$ represents the contribution from higher Landau levels:
\begin{equation}
    V_{\text{nonH}}=\sum_{m=2}^{\infty}u(-,m,\vec{0})+\sum_{m=0}^{\infty}u(+,m,\vec{0}).
\end{equation}
It can be shown that in the limit $H\rightarrow 0$, only this part yields a non-zero contribution to the potential, reducing to the effective potential found in reference \cite{Chen:2022smf} (see Appendix~\ref{C}). Therefore, it is numerically impossible to achieve a perfectly vanishing $H$. To ensure convergence, the summation over Landau levels must be carried out up to an order of magnitude of approximately $O(H^{-1})$. In our calculation, we set a lower limit of the scaled variable $\beta\sqrt{gH}=0.01$, at which point $V_{\text{nonH}}$ essentially coincides with the exact value, and the chromomagnetic field at this setting is regarded as vanishing. In the subsequent discussion, we will identify this as the chromomagnetic-suppressing potential.

Furthermore, the above expressions admit an alternative representation in series form. Using the series expansion $\ln(1-z)=-\sum_{n=1}^{\infty}z^n/n$ for $|z|\leq1$ and the following integral identities \cite{Meisinger:2002ji}:
\begin{gather}
    K_1(n\beta m)=\frac{1}{2m}\int_{-\infty}^{+\infty}dke^{-n\beta\sqrt{k^2+m^2}},\\
\lim_{\epsilon\rightarrow0^+}\operatorname{Re}\int_{0}^{+\infty} \frac{dk}{2\pi} e^{-n \sqrt{k^2 - k_c^2 - i\epsilon}}=-\frac{k_c}{4}Y_1(nk_c),
\end{gather}
we obtain the series expressions:
\begin{widetext}
\begin{gather}
    u(-,0,\vec{0})=\frac{(gH)^\frac{3}{2}}{\pi^2\beta} \sum_{n=1}^\infty \frac{1}{n}
        \frac{\pi}{2} Y_1(n\beta\sqrt{gH})
      \cos n( \phi + \tilde{\Omega}),\\
    u(-,1,\vec{0})=- \frac{(gH)^\frac{3}{2}}{\pi^2\beta} \sum_{n=1}^\infty \frac{1}{n}
       K_1(n\beta\sqrt{gH})
      \cos n( \phi + \tilde{\Omega}), \\
    V_{\text{nonH}}=- 2\frac{(gH)^\frac{3}{2}}{\pi^2\beta}
      \sum_{n=1}^\infty \sum_{m=0}^\infty \frac{1}{n} \sqrt{2m + 3}
      K_1(n\beta\sqrt{gH(2m + 3)}) \cos n \phi \cos n\tilde{\Omega}.
\end{gather}
Numerical calculations indicate that the series representation for $V_{\text{nonH}}$ converges faster, while the opposite is true for $u(-,0,\vec{0})$. Consequently, in the final calculation of the total potential, we employ a hybrid approach, using the integral representation for some terms and the series representation for others:
\begin{align}
V(r=0) &= \frac{11g^2H^2}{48\pi^2} \ln \left( \frac{gH}{ \Lambda^2} \right)
      +\frac{gH}{2\pi \beta}\sum_{s=\pm}\int_{-\infty}^{+\infty}\frac{dk}{2\pi} \ln \bigg|1-2\cos{(\phi+\tilde{\Omega})}e^{-\beta\sqrt{k^2+sgH}}+e^{-2\beta\sqrt{k^2+sgH}}\bigg| \notag \\
    &\quad - 2\frac{(gH)^\frac{3}{2}}{\pi^2\beta}
      \sum_{n=1}^\infty\frac{1}{n} \cos n \phi \cos n\tilde{\Omega}  \sum_{ m=0}^\infty \sqrt{2 m + 3}
      K_1(n\beta\sqrt{(2 m + 3)gH}) .
\end{align}
\end{widetext}
As is evident, under the condition that $H \neq 0$ and $\tilde{\Omega} \neq 0$, the presence of the Lowest Landau Level (LLL) leads to a thermodynamic potential which no longer respects the symmetry $V(\phi) = V(2\pi - \phi)$. Consequently, the $\mathbb{Z}(2)$ center symmetry undergoes explicit breaking. This explicit breaking is likely to exert a non-trivial influence on the characteristics of the perturbative confinement phase transition induced by $\Omega$. In all subsequent calculations of the effective potential, we will adopt the dimensionless form $\beta^4 V$ for computation and analysis. The associated gap equations are given by
\begin{equation}
    \frac{\partial V}{\partial H} = \frac{\partial V}{\partial\phi} = 0.
\end{equation}

\section{Effects of Temperature and Imaginary Angular Velocity}
\label{sec:3}
Within the chromomagnetic background field, we define the effective potential for the Polyakov loop as:
\begin{equation}
    V(\phi)=V( \langle H \rangle,\phi),
\end{equation}
where $\langle H \rangle$ satisfies:
\begin{equation}
    \partial_HV(H,\phi)\Big|_{H=\langle H \rangle}=0.
\end{equation}

As demonstrated in reference \cite{Meisinger:2002ji}, there exists a complex dependence between the chromomagnetic condensation $H$ and the Polyakov loop $\phi$. Notably, for $T<0.2\Lambda$, the Savvidy model exhibits intriguing oscillations between confined and deconfined phases as the temperature varies. When $T>0.2\Lambda$, this oscillatory behavior disappears. Within the temperature interval $\Lambda^{-1}T \in (0.20,0.72)$, the minimum of $V(\phi)$ resides at the confining, non-trivial Polyakov loop value $\phi=\pi$. As the temperature exceeds the upper threshold, it discontinuously jumps to the deconfining, trivial Polyakov loop values $\phi=0$ and $2\pi$. We now investigate the evolution of the Polyakov loop potential with increasing $T$ to observe this behavior with $\Omega=0$.

\begin{figure*}[t]
    \centering
    % 第一行：图片 1 和图片 2
    \begin{subfigure}[b]{0.45\textwidth}
        \includegraphics[width=\textwidth]{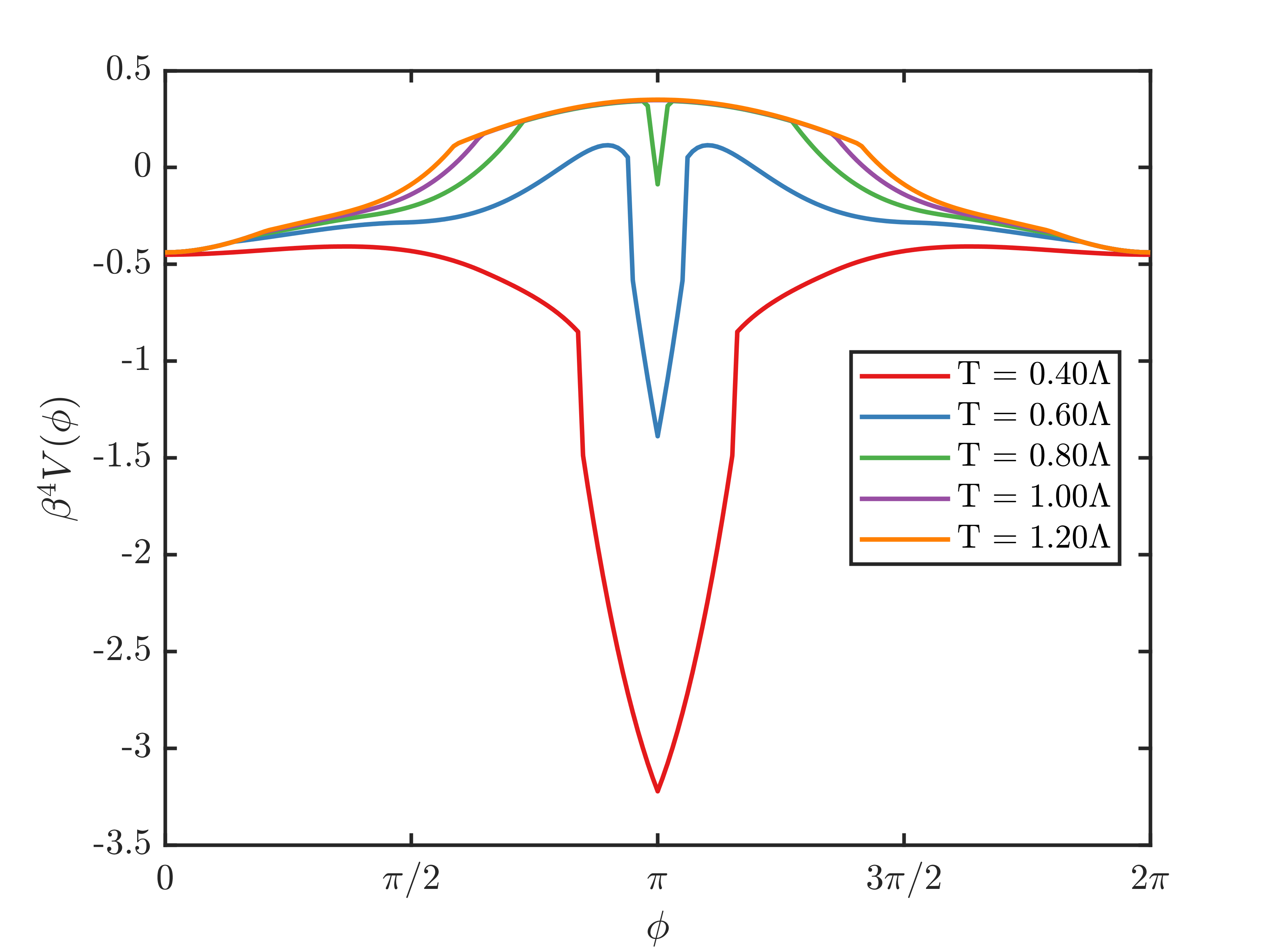}
        \phantomcaption
    \end{subfigure}
    \hspace{-0.5cm} % 负间距缩小水平间隔
    \begin{subfigure}[b]{0.45\textwidth}
        \includegraphics[width=\textwidth]{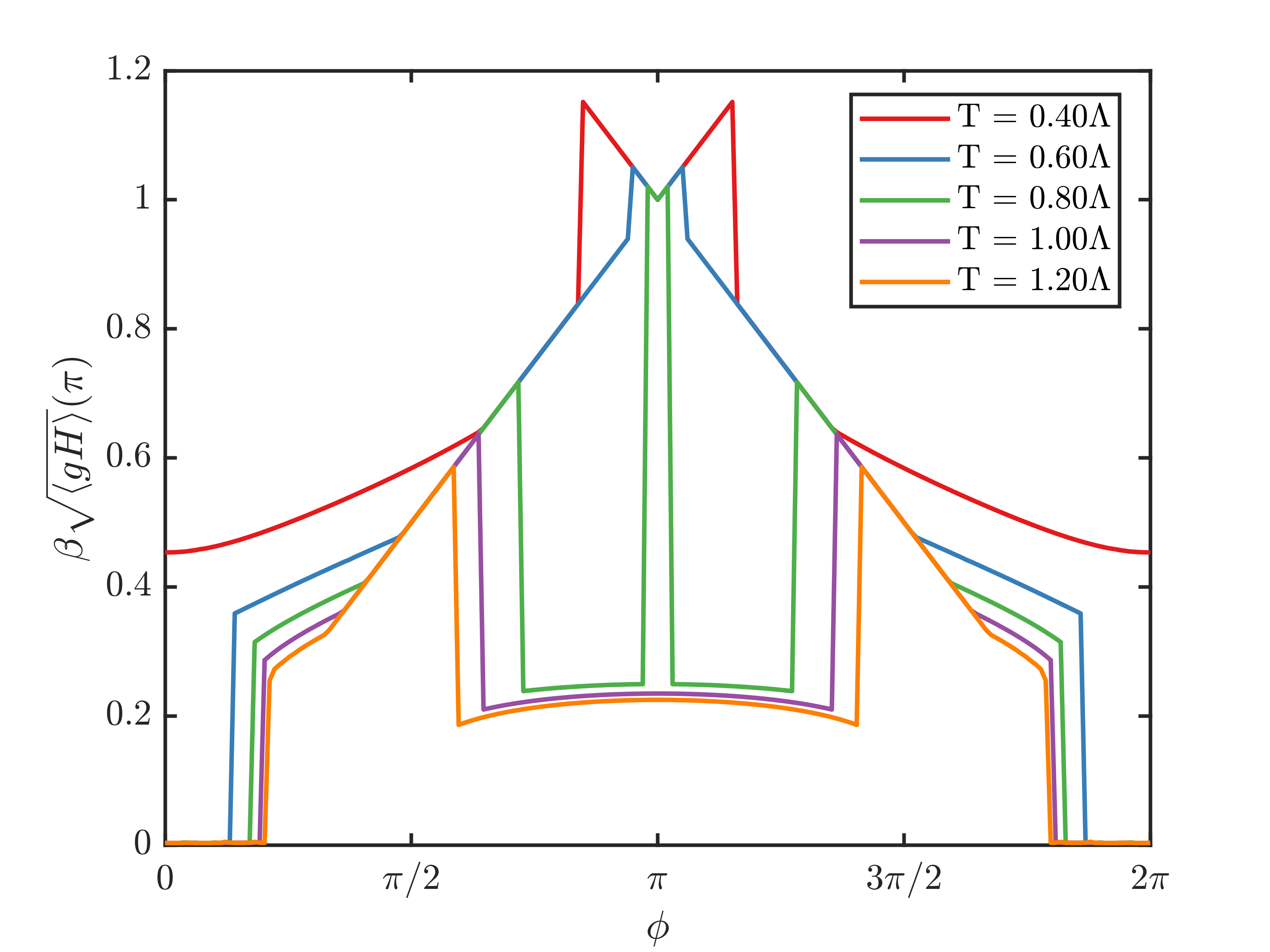}
        \phantomcaption
    \end{subfigure}
 % \vspace*{-0.5cm}
 \caption{Evolution of the dimensionless Polyakov loop potential $\beta^4 V$ (left) and the corresponding scaled variable $\beta \sqrt{\langle gH \rangle}$ (right) as functions of $\phi$ at several temperatures, for $\Omega = 0$ and $r=0$.}
 \label{fig1}
\end{figure*}
Fig.~\ref{fig1} shows the evolution of the dimensionless Polyakov loop potential (scaled by $T^4$) in terms of $\phi$ with increasing temperature at $\Omega=0$ and $r=0$, alongside the corresponding scaled variable $\beta \sqrt{\langle gH \rangle}$. At any temperature, the potential is symmetric about $\phi=\pi$, reflecting the $\mathbb{Z}(2)$ center symmetry. The dependence of $\beta \sqrt{\langle gH \rangle}$ on $\phi$ reveals the intricate entanglement between chromomagnetic condensation and the Polyakov loop. At low temperatures, the Polyakov loop potential features two types of minima: shallow ones located at $\phi=0, 2\pi$ and a sharp one at $\phi=\pi$. Thermal effects suppress the minimum at $\phi=\pi$. Upon reaching a certain threshold temperature, the global minimum discontinuously jumps to $\phi=0, 2\pi$, and $\beta \sqrt{\langle gH\rangle}$ simultaneously jumps from $\pi$ (which is determined by the LLL \cite{Meisinger:2002ji}) to $0$. This suggests a first-order deconfinement phase transition. It is noteworthy, however, that lattice results have long established that the $\mathrm{SU}(2)$ deconfinement transition belongs to the three-dimensional Ising universality class \cite{Engels:1994xj}, corresponding to a second-order transition. The discrepancy arises because our one-loop calculation is not applicable in the low-temperature, non-perturbative regime. In subsequent calculations, we set a lower temperature limit $T_\text{pert}=10\Lambda$ to ensure reliability, treating the region $T<T_\text{pert}$ as the  Strong Coupling Transition Region.

As shown in references \cite{Meisinger:2002ji}, the contributions to the effective potential from both the LLL and the pure vacuum energy density each exhibit their minimum at a non-vanishing chromomagnetic field.  leading us to identify the  chromomagnetic-favoring potential  $V_H$:
\begin{align}
    V_H &=\frac{11g^2H^2}{48\pi^2} \ln \left( \frac{gH}{ \Lambda^2} \right)
      +\frac{gH}{2\pi\beta}\sum_{s=\pm}\int_{-\infty}^{+\infty}\frac{dk_z}{2\pi}\notag\\
      &\ln \bigg|1-2\cos{(\phi+\tilde{\Omega})}e^{-\beta\sqrt{k_z^2+sgH}}+e^{-2\beta\sqrt{k_z^2+sgH}}\bigg|.
\end{align}
Calculations for the high-temperature part in Figure~\ref{fig1}  reveal that chromomagnetic condensation is strongly suppressed at the minimum, effectively vanishing. This state persists in the region $T>T_\text{pert}$. A rough analysis based on the dimensionless vacuum energy (scaled by $T^4$) shows that thermal effects suppress chromomagnetic condensation, shifting the minimum of $V_H$ towards smaller $\beta\sqrt{gH}$ and higher potential energy as temperature increases. Consequently, the overall potential tends to disfavor the generation of the scaled chromomagnetic condensation.

\begin{figure*}[t]
    \centering
    \includegraphics[width=18cm]{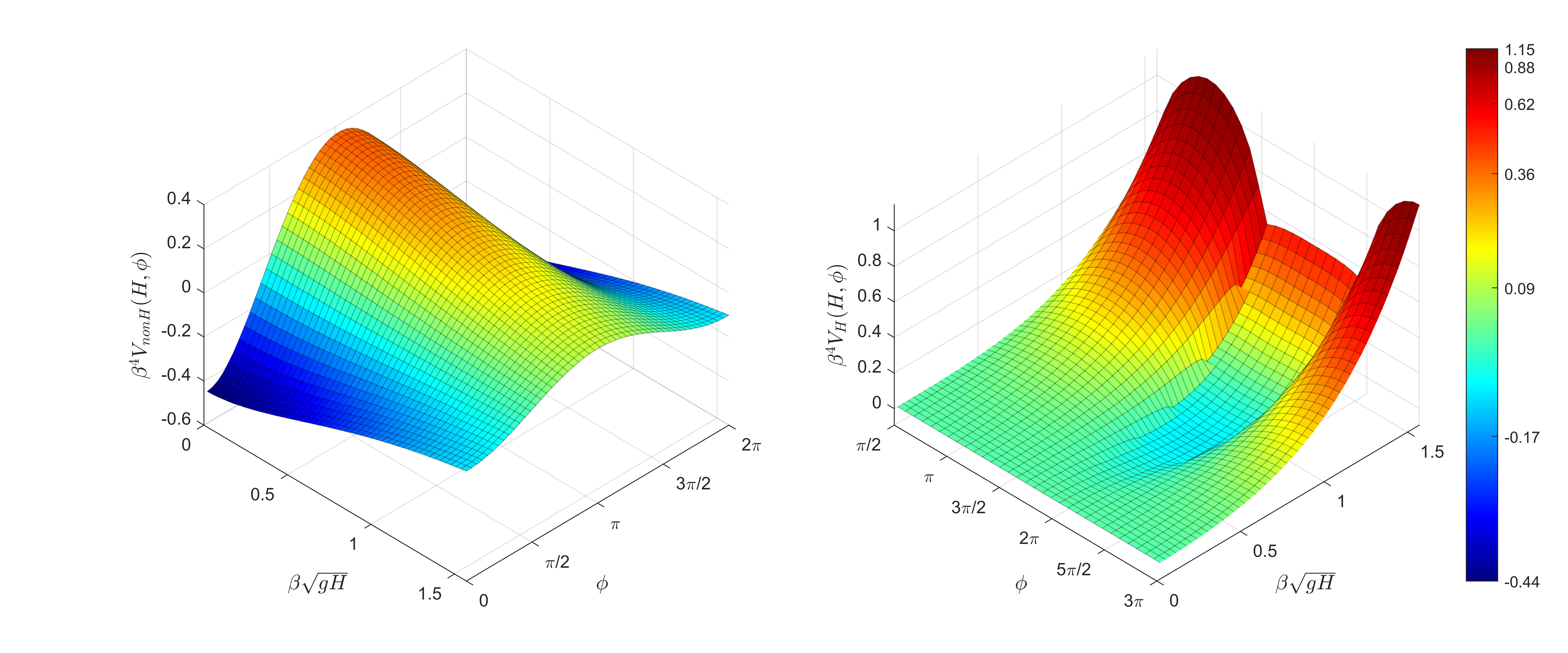}
    \caption{Three-dimensional plots of the dimensionless potentials $\beta^4 V_{\text{nonH}}$ (left) and $\beta^4 V_H$ (right) as functions of $\phi$ and $\beta\sqrt{gH}$ for $\tilde{\Omega}=0$.}
    \label{fig2}
\end{figure*}

Figure~\ref{fig2} shows the chromomagnetic-favoring potential $V_H(H,\phi)$ and the chromomagnetic-suppressing potential $V_\text{nonH}(H,\phi)$ at $\Omega=0$ and $T=10\Lambda$. $V_\text{nonH}$ is independent of temperature. The left panel clearly shows that the global minimum is located at $\phi=0,2\pi$ with $H\rightarrow0$. $V_\text{nonH}$ exhibits a suppressing effect on chromomagnetism: as $H$ increases, $|V_\text{nonH}|$ becomes smaller, meaning the potential energy at $\phi=0,2\pi$ rises, bringing the potential contour closer to the $V=0$ plane. According to Appendix~\ref{C}, we can analyze this roughly. $V_\text{nonH}$ satisfies the expression:
\begin{equation}
    V_\text{nonH}=\Delta x\sum_{n=1}^{\infty}f\left(x_n\right),
\end{equation}
which represents an integral expression with step size $\Delta x = \beta^2 gH$, variable $x_n=n\beta^2gH$, and integration interval $[0, +\infty)$, where the integrand $f$ is monotonic and convergent in $x$. A larger step size leads to a smaller absolute value of the integral. The right panel shows that the minimum of $V_H$ is located in the region with $\phi=0,2\pi$ and $H \neq 0$. However, this minimum is at a significantly higher energy level compared to $V_\text{nonH}$ and holds no advantage. When the imaginary angular velocity is zero, the physical ground state in the perturbative regime is entirely dominated by $V_\text{nonH}$, and chromomagnetism is suppressed.

However, the situation changes as $\tilde{\Omega}$ increases, because the effect of the imaginary angular velocity on $V_H$ and $V_\text{nonH}$ is markedly different. Based on the overall symmetry of the potential:
\begin{equation}
    V(H,\phi,\tilde{\Omega})=V(H,3\pi-\phi,\pi-\tilde{\Omega}),
\end{equation}
our subsequent analysis of the imaginary angular velocity effect can be confined to the interval $\tilde{\Omega}\in [0,\pi/2]$. The physical minimum $(H,\phi)$ satisfies:
\begin{equation}
    \left( H ,\phi \right)\Big|_{\pi -\tilde{\Omega}}=\left( H ,3\pi- \phi \right)\Big|_{\tilde{\Omega}}.
    \label{symmetry}
\end{equation}
First, from the expression for $V_H$, it is straightforward to see that the imaginary angular velocity shifts the entire potential $V_H(H,\phi)$ negatively along the $\phi$ direction. The location and depth of its minimum remain unchanged, but the corresponding $\phi$ value shifts from $2\pi$ to $2\pi-\tilde{\Omega}$.

\begin{figure}[htbp]
    \centering
    \includegraphics[width=8.6cm]{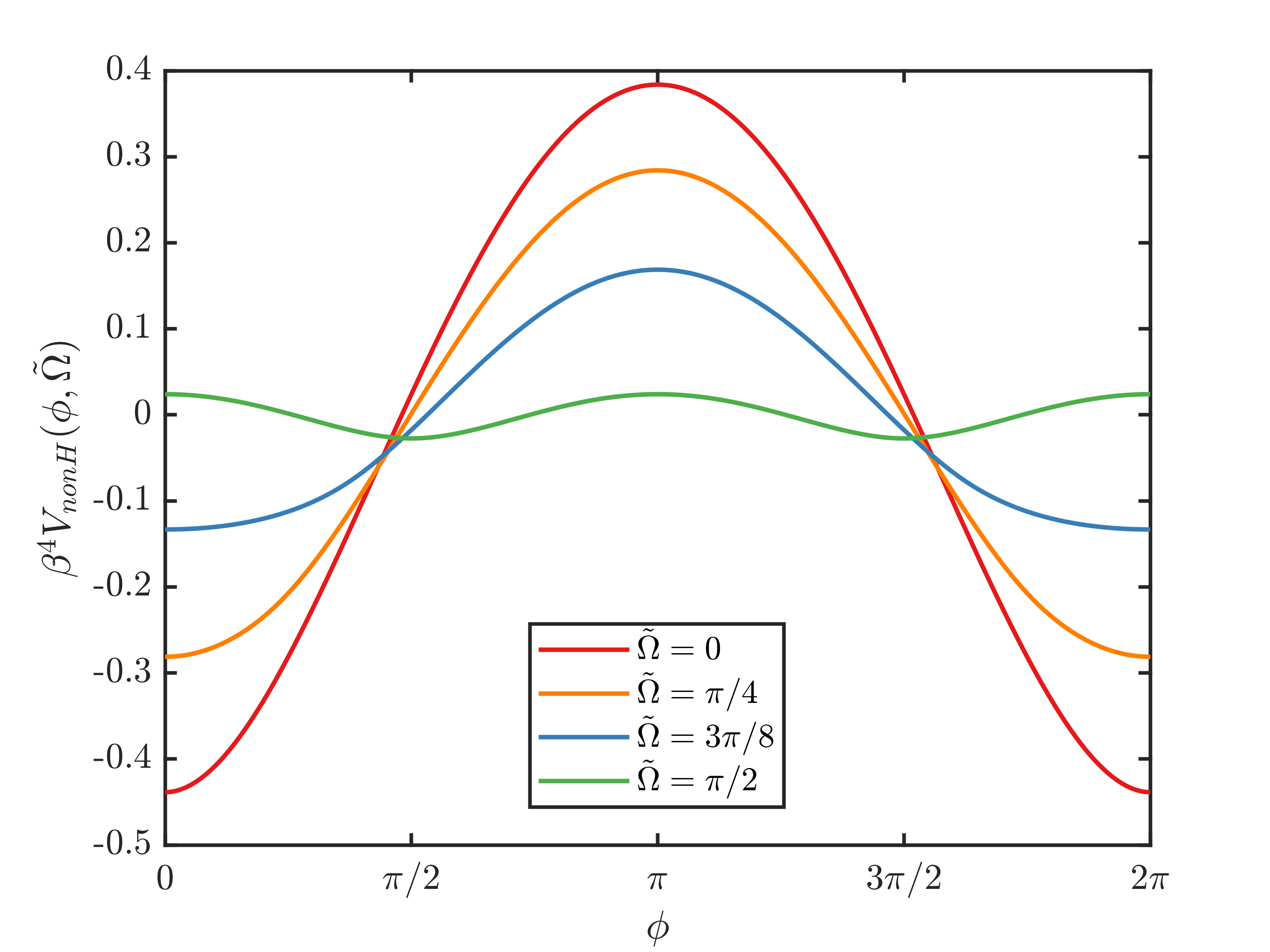}
    \caption{Evolution of the dimensionless $\beta^4 V_{\text{nonH}}$ chromomagnetic-suppressing potential  with $\phi$ for $\tilde{\Omega}=0, \pi/4, 3\pi/8, \pi/2$ at $H\rightarrow0$ and $r=0$.}
    \label{fig3}
\end{figure}
 Fig.~\ref{fig3} shows the evolution of the chromomagnetic-suppressing potential in terms of $\phi$ with increasing $\tilde{\Omega}$ at $H\rightarrow0$ and $r=0$. It is evident that the potential energy at the physical minimum rises rapidly with increasing $\tilde{\Omega}$. This implies that although the minimum of $V_\text{nonH}$ still corresponds to $H\rightarrow0$, its ability to suppress chromomagnetism weakens significantly. Compared to the translational effect of $\tilde{\Omega}$ on $V_H$, $V_\text{nonH}$ gradually loses its competitive advantage. Eventually, between the minima of these two potential components, a new global minimum for the total potential emerges. It can be anticipated that as the imaginary angular velocity increases to a critical value $\tilde{\Omega}_{c}\in (0,\pi/2)$, the physical ground state will discontinuously jump from $(H, \phi)=(0,2\pi)$ to a point with $H \neq 0$ and $\phi \in(2\pi-\tilde{\Omega}_c,2\pi)$. Using the symmetry relation in Eq.~(\ref{symmetry}), a symmetric jump from a point with $H \neq 0$ and $\phi \in(\pi,\pi+\tilde{\Omega}_c)$ to the confining phase at $(H, \phi)=(0,\pi)$ occurs at $\tilde{\Omega}=\pi-\tilde{\Omega}_c$. Thus, the perturbative confinement phase is reached via a first-order transition, and  two first-order phase transitions are expected during the emergence of the perturbative confinement phase.

\section{Perturbative confinement phase transition and phase diagram}
\label{sec:4}
In the previous section, we made predictions about the nature of the perturbative confinement phase transition based on the effects of $V_H$ and $V_\text{nonH}$ on chromomagnetism, as well as the reshaped competition between them under rotation. In this section, we perform detailed calculations to explore the phase diagram and properties of the perturbative confinement phase transition.

\begin{figure*}[t]
    \centering
    % 第一行：图片 1 和图片 2
    \begin{subfigure}[b]{0.45\textwidth}
        \includegraphics[width=\textwidth]{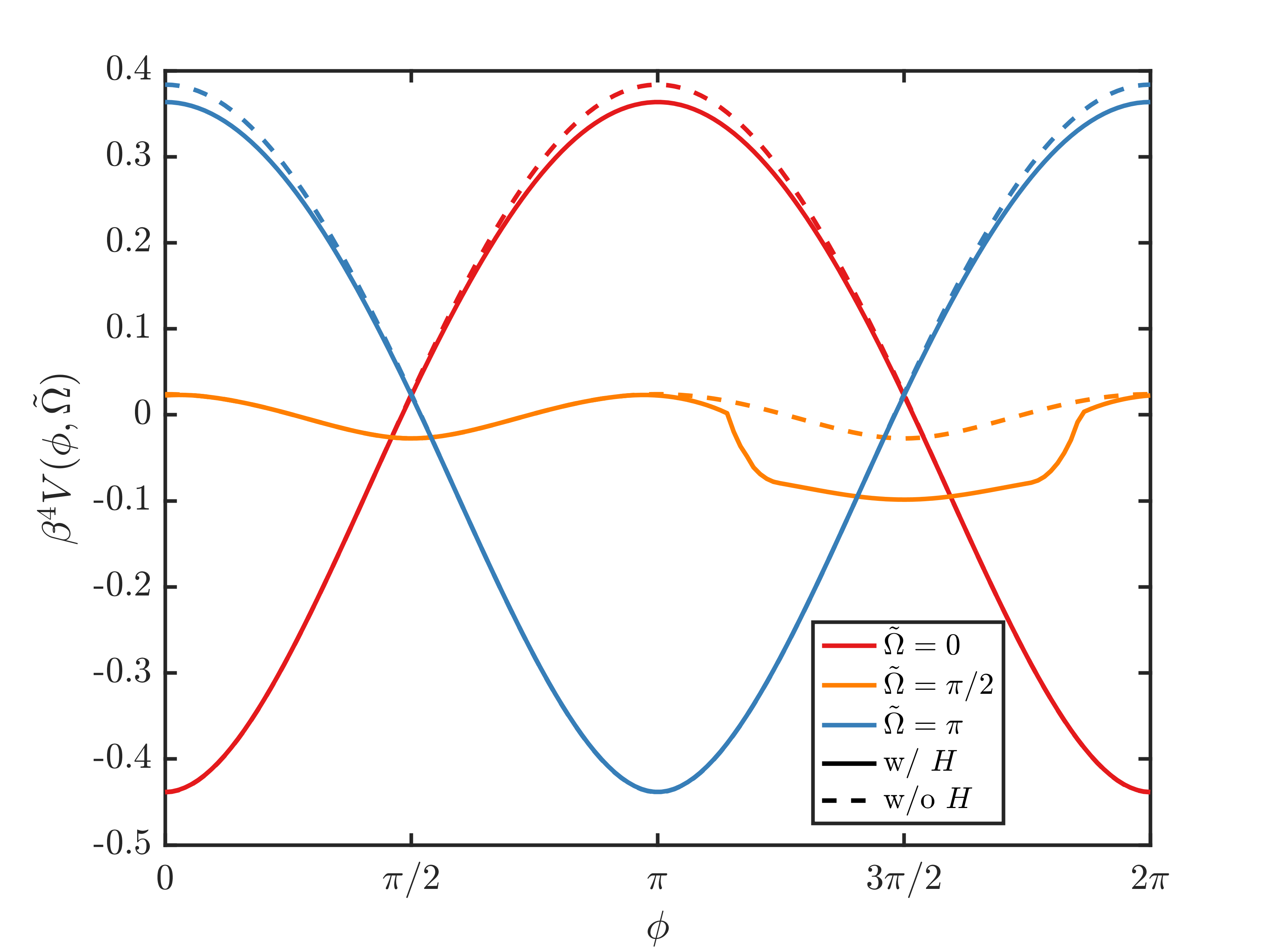}
        \phantomcaption
    \end{subfigure}
    \hspace{-0.5cm} % 负间距缩小水平间隔
    \begin{subfigure}[b]{0.45\textwidth}
        \includegraphics[width=\textwidth]{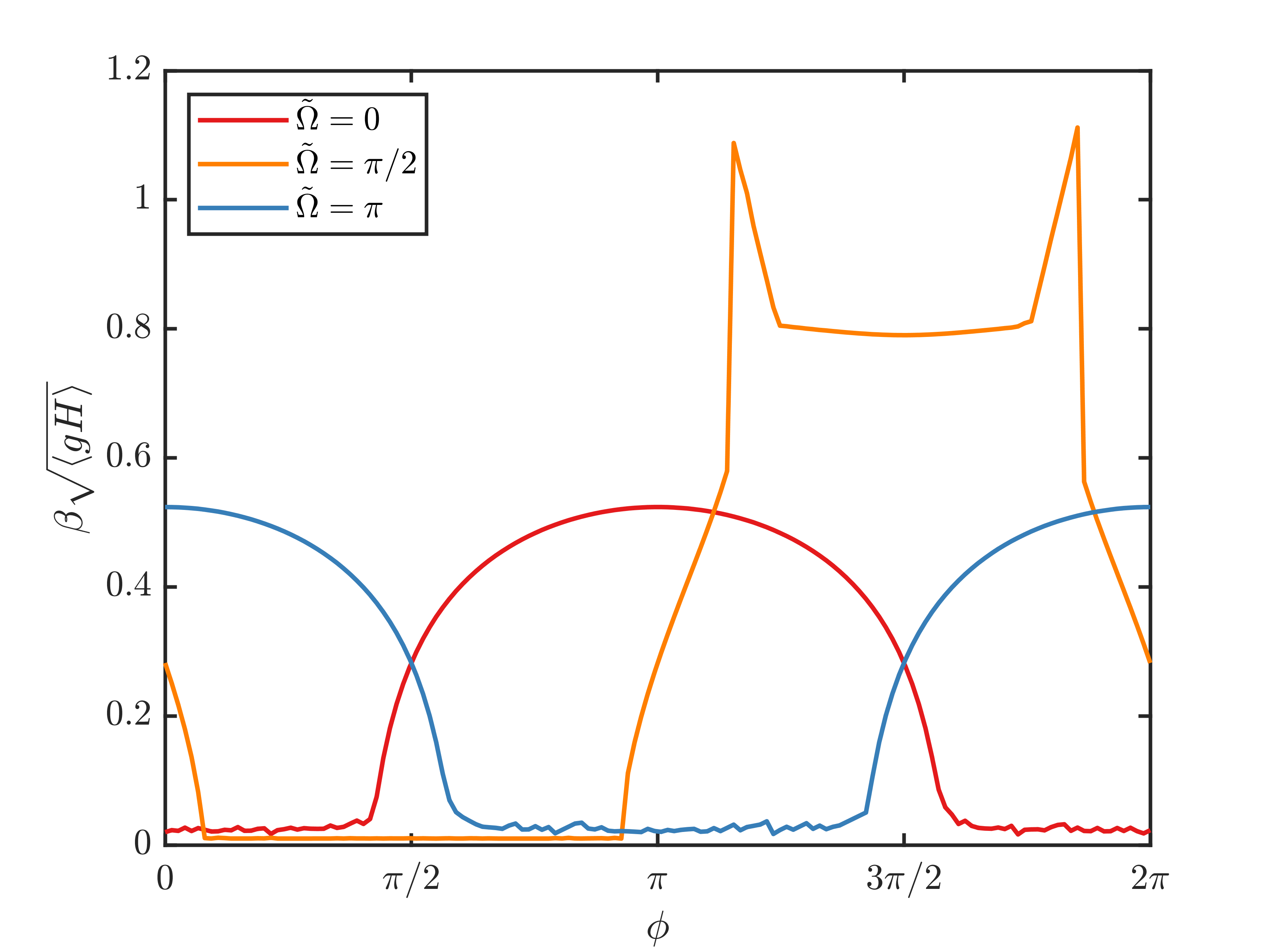}
        \phantomcaption
    \end{subfigure}
 % \vspace*{-0.5cm}
 \caption{Evolution of the dimensionless Polyakov loop potential $\beta^4 V$ (left) and the corresponding scaled variable $\beta \sqrt{\langle gH \rangle}$ (right) with $\phi$ for several values of $\tilde{\Omega}$ at $T=10\Lambda$ and $r=0$.}
 \label{fig4}
\end{figure*}
 Fig.~\ref{fig4} shows the evolution of the Polyakov loop potential (with respect to $\phi$) with increasing $\tilde{\Omega}$ at $r=0$ and $T=10\Lambda$, both in the presence of a chromomagnetic background field (solid lines) and in its absence (dashed lines), along with the corresponding $\beta \sqrt{\langle gH \rangle}$. First, we can clearly see that the potential curve with a chromomagnetic background is always lower than that without. In the left panel, the regions where the two curves nearly coincide correspond to very low values of $\beta \sqrt{\langle gH \rangle}$ in the right panel, which fluctuate at very low magnitudes and can be considered as vanishing. In regions where the potential curves deviate significantly, the chromomagnetic condensate is induced, and the potential reaches a lower value. The red curve in Fig.~\ref{fig4} for $\tilde{\Omega}=0$ reproduces the spontaneously broken center symmetry with minima located at $\phi=0$ and $2\pi$. The orange curve for $\tilde{\Omega}=\pi/2$ has its minimum shifted to $\beta\sqrt{\langle gH \rangle}=0.8$ and $\phi=3\pi/2$, indicating that the imaginary angular velocity induces a chromomagnetic condensate and further leads to explicit breaking of the center symmetry. The blue curve for $\tilde{\Omega}=\pi$ shows that as $\tilde{\Omega}$ increases, the potential minima deviate from the deconfined vacua to the confined vacuum at $\phi=\pi$.

\begin{figure*}[t]
    \centering
    % 第一行：图片 1 和图片 2
    \begin{subfigure}[b]{0.45\textwidth}
        \includegraphics[width=\textwidth]{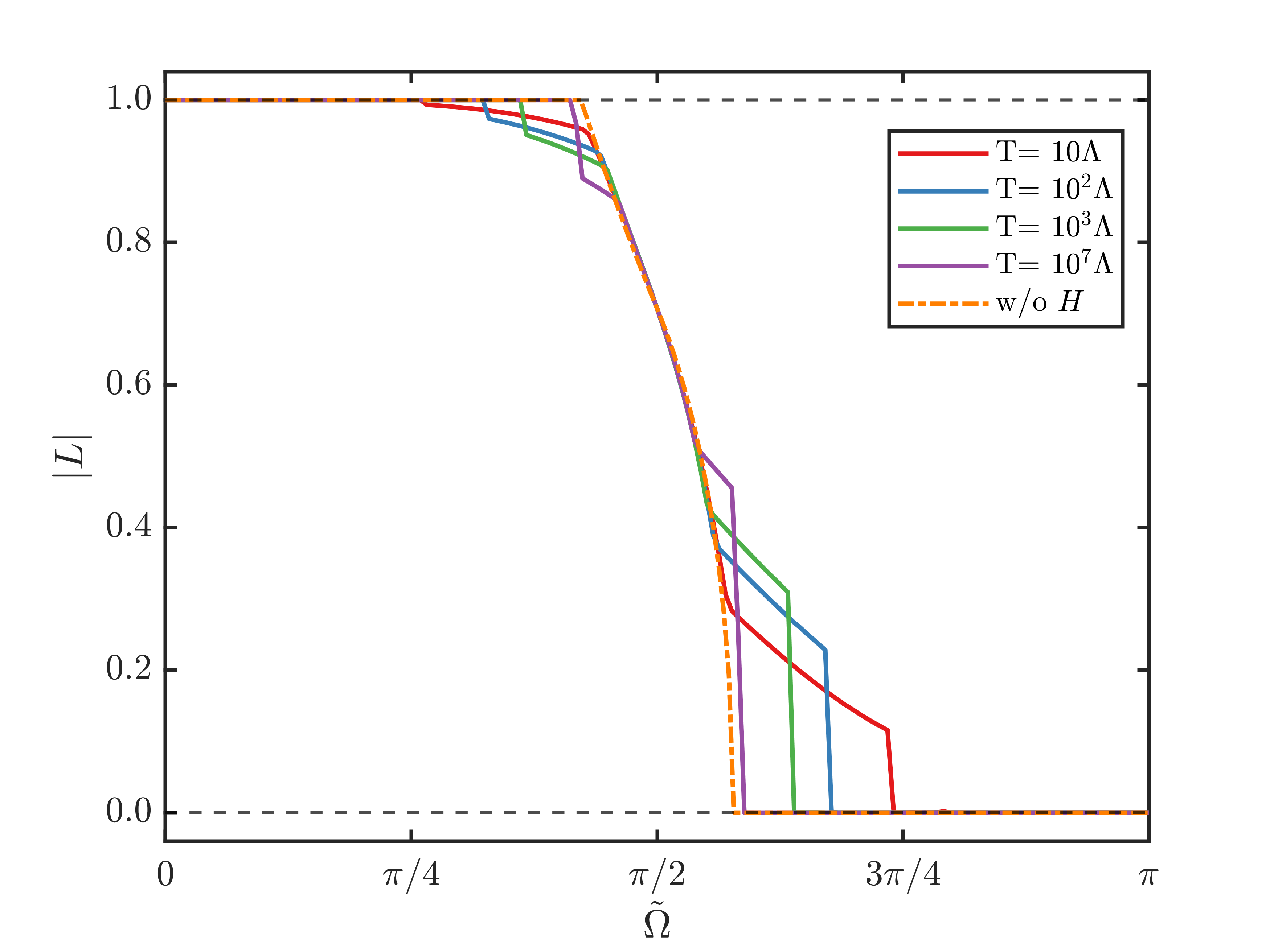}
        \phantomcaption
    \end{subfigure}
    \hspace{-0.5cm} % 负间距缩小水平间隔
    \begin{subfigure}[b]{0.45\textwidth}
        \includegraphics[width=\textwidth]{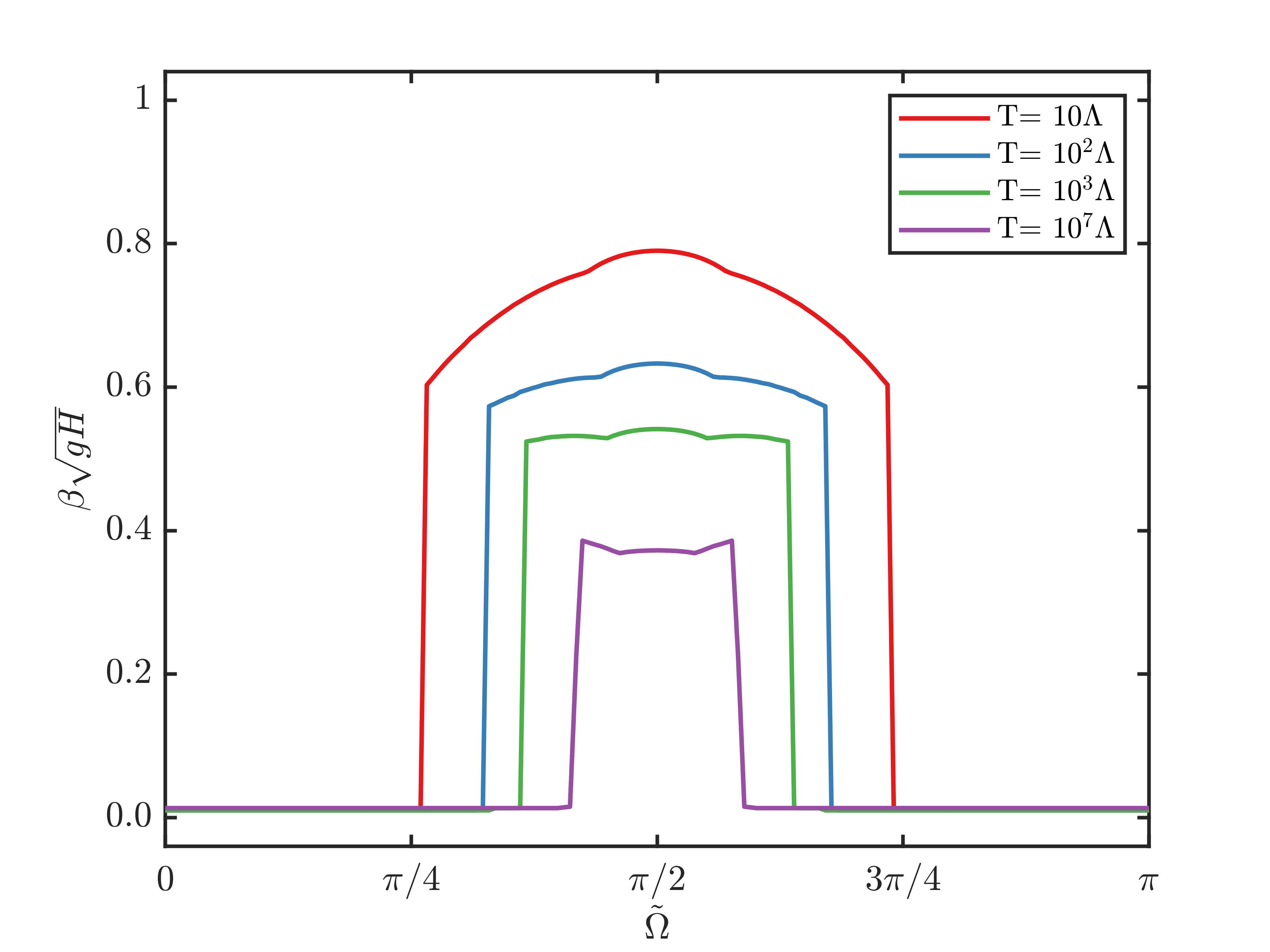}
        \phantomcaption
    \end{subfigure}
 % \vspace*{-0.5cm}
 \caption{The expectation value of the fundamental Polyakov loop $|L|$ (left) and the scaled chromomagnetic condensate $\beta\sqrt{gH}$ (right) as functions of $\tilde{\Omega}$ at different temperatures.}
 \label{fig5}
\end{figure*}
We can visualize this phase transition by plotting the expectation value of the fundamental Polyakov loop $|L|$ and the scaled chromomagnetic condensate $\beta\sqrt{gH}$ as functions of $\tilde{\Omega}$ at different temperatures, as shown in Fig.~\ref{fig5}. In the left panel, we can see that at different temperatures, $|L|$ undergoes several nontrivial changes during the emergence of the perturbative confinement phase. We need to analyze this step by step. First, the most obvious nontrivial change occurs at imaginary angular velocities symmetric about $\tilde{\Omega}=\pi/2$, where two first-order phase transitions take place, corresponding to jumps in the chromomagnetic field $H$ and the Polyakov loop. This is entirely consistent with our predictions in the previous section. Moreover, due to the suppressing effect of temperature on $V_H$, the scaled chromomagnetic condensate  $\beta\sqrt{gH}$ induced by $\tilde{\Omega}$ becomes smaller at higher temperatures. For $\tilde{\Omega}\in (0,\pi/2)$, a larger $\tilde{\Omega}$ is required to achieve greater suppression of $V_\text{nonH}$ in order to induce the chromomagnetic condensate. Correspondingly, the critical $\tilde{\Omega}_c\in(\pi/2,\pi)$ for the perturbative confinement transition decreases with increasing temperature, gradually approaching the critical value $\tilde{\Omega}_c=\pi/\sqrt{3}$ for the case without a chromomagnetic background field, indicated by the dashed line in the figure. Second, after the first-order phase transition in the interval $\tilde{\Omega}\in (0,\pi/2)$, the global minimum of the total potential lies between the minima of $V_H$ and $V_\text{nonH}$ at $(H,\phi)=(0,2\pi)$ and $(H\neq 0,2\pi-\tilde{\Omega})$. However, we know that when $\tilde{\Omega}$ becomes even larger, the minimum of $V_H$ rapidly shifts from $\phi=2\pi$ to $\phi=\pi$. Influenced by this effect, the evolution of $\phi$ with $\tilde{\Omega}$ also exhibits a turning, descending more rapidly with increasing $\tilde{\Omega}$, and the evolution of the chromomagnetic condensate is also affected to some extent.

\begin{figure}[htbp]
    \centering
    \includegraphics[width=8.6cm]{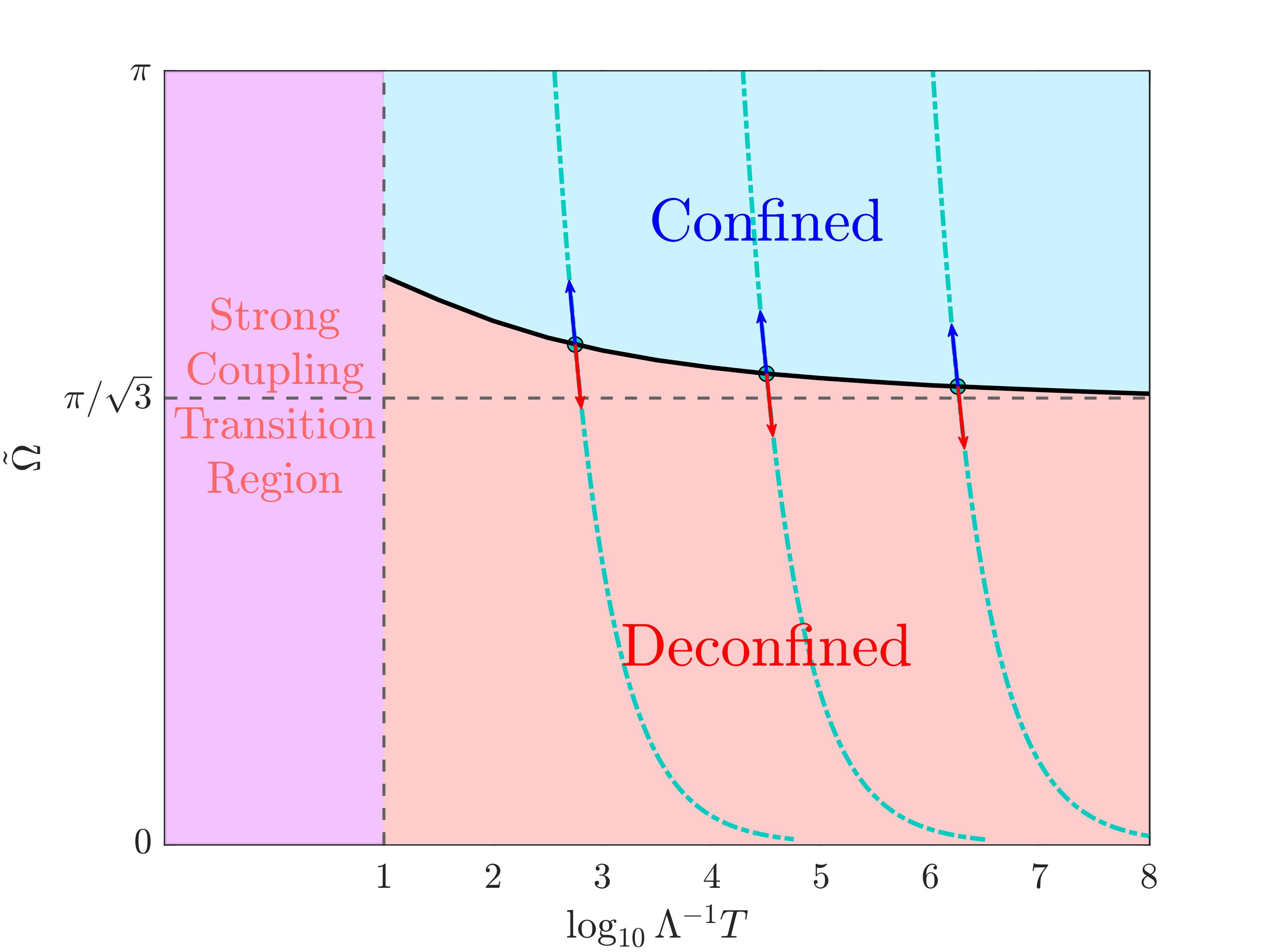}
    \caption{Phase diagram for the perturbative confinement transition in the $T$-$\tilde{\Omega}$ plane.}
    \label{fig6}
\end{figure}

Based on the above observations, it is tempting to envision a new phase diagram for the perturbative confinement phase transition on the $T$-$\tilde{\Omega}$ plane, as shown in Fig.~\ref{fig6}. To better illustrate the temperature dependence of the phase boundary and its asymptotic behavior at high temperatures, we take the horizontal coordinate as $\log_{10}{\Lambda^{-1}T}$ and the vertical coordinate as $\tilde{\Omega}$. By analyzing the curve for the first-order phase transition boundary, we can see that as temperature increases, the critical $\tilde{\Omega}_c$ decreases and asymptotically approaches the result without chromomagnetism, $\tilde{\Omega}_c=\pi/\sqrt{3}$. Moreover, within the $T$-$\tilde{\Omega}$ region where the phase boundary curve is calculated, the critical temperature $T_c$ decreases with increasing $\tilde{\Omega}$. Furthermore, for a fixed $\tilde{\Omega}$, looking from either side of any point on the phase boundary, increasing temperature causes the system to transition from the deconfined phase to the perturbatively confined phase. This contradicts the conventional view that increasing temperature tends to drive the system towards deconfinement. To isolate the pure temperature effect, we have marked three constant-$\Omega$ lines in the figure (from left to right, $\Omega$ increases). The red arrows indicate the direction of increasing temperature, while the blue arrows indicate decreasing temperature. It can be seen that at fixed $\Omega$, the temperature effect still tends to drive the system towards deconfinement, maintaining the conventional picture. Moreover, $T_c$ increases monotonically with $\Omega$.

It is worth noting that the upper limit of temperature in our calculations here is $T=10^8\Lambda$. Although asymptotic freedom suggests that higher temperatures are more favorable for perturbative calculations, extremely high temperatures may lose practical physical meaning due to the involvement of other higher-energy physical processes.

\begin{figure}[htbp]
    \centering
    \includegraphics[width=8.6cm]{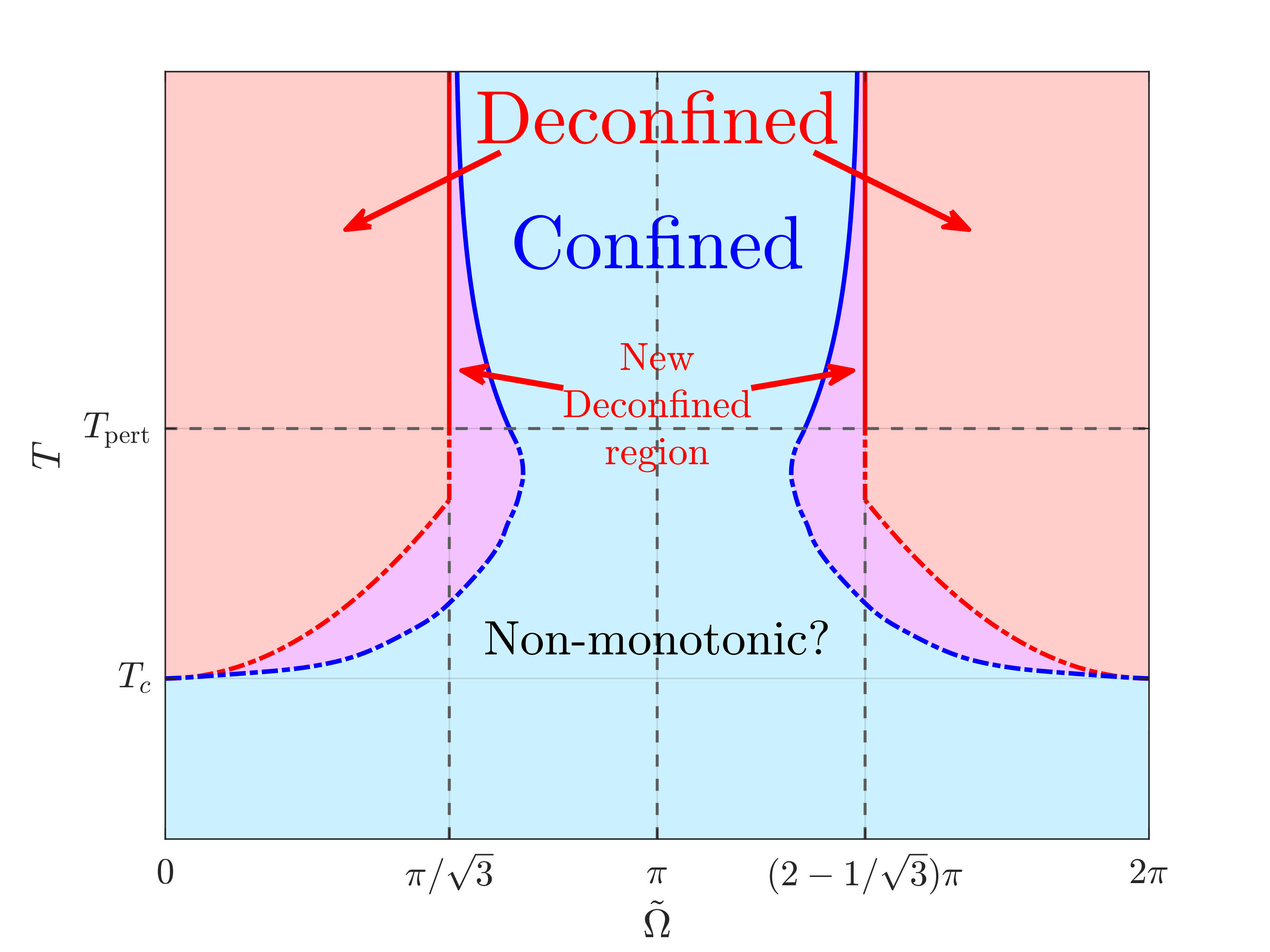}
    \caption{Conjectured global phase diagram on the $\tilde{\Omega}$-$T$ plane near the rotation axis ($r=0$) for $\mathrm{SU}(2)$ Yang-Mills theory, with curves representing phase transitions.}
    \label{fig7}
\end{figure}

The speculated phase diagram on the entire $\tilde{\Omega}$-$T$ plane in reference \cite{Chen:2022smf} has attracted widespread attention and discussion. Through the above research, we find that the introduction of a chromomagnetic background leads to a richer phase structure for the perturbative confinement transition. For comparison, we also plot the phase diagram in Fig.~\ref{fig7} to illustrate our main conclusions. The most obvious change after introducing the chromomagnetic background is that we have identified and confirmed that a larger region at high temperatures belongs to the deconfined region (purple area in the figure). Additionally, the phase boundary curve (blue line) in this region exhibits significant temperature dependence, which differs from previous conclusions (red line). Regarding the order of the phase transition, considering the excitation effect of $\tilde{\Omega}$ on the chromomagnetic condensate and the explicit breaking of $\mathbb{Z}(2)$ symmetry, our results suggest that the perturbative confinement transition is first-order rather than second-order. In the low-temperature non-perturbative region, as analyzed in references \cite{Chen:2022smf,Braun:2007bx}, the Kugo-Ojima-Gribov-Zwanziger (KOGZ) mechanism \cite{Gribov:1977wm,Kugo:1979gm,Zwanziger:1993dh} causes the ghost fields, which cancel unphysical polarization modes at high temperatures, to make the system increasingly confining uniformly for any $\tilde{\Omega}$. However, our perturbative calculations indicate that before entering the non-perturbative temperature region, there might still be an interval where the critical $\tilde{\Omega}_c$ increases with decreasing temperature. From an overall perspective, however, there should exist a general trend in the non-perturbative regime where the critical value $\tilde{\Omega}_c$ increases with rising temperature. From the standpoint of phase boundary continuity, it is entirely possible that the phase boundary exhibits non-monotonic behavior within the non-perturbative regime, indicating the presence of complex non-perturbative effects. This would, of course, significantly impact the adiabatic continuity evolution path without a phase transition—connecting the perturbatively confined phase and the conventional confined phase—as mentioned in reference \cite{Chen:2022smf}. Last but not least, given that the non-perturbative deconfinement transition in $\mathrm{SU}(2)$ theory is of second order \cite{Engels:1994xj}, a critical end point (CEP) may exist, connecting the first-order perturbative confinement transition and the second-order non-perturbative deconfinement transition. 
\section{Analytical continuation to real rotation.}
\label{sec:5}
As mentioned earlier, due to the sign problem in lattice calculations, in order to obtain results for real angular velocity corresponding to physical reality, the lattice method extrapolates the dependence of related physical quantities (such as $T_c$ on $\Omega$) via the relation $\Omega^2=-\omega^2$. To be precise, they must obtain the dependence on $\Omega^2$ to perform a smooth analytic continuation. For instance, the Tolman-Ehrenfest effect verified in reference \cite{Chernodub:2022veq} also depends on $\Omega^2$.

In reference \cite{Chen:2022smf}, it was found that the local effective potential fully satisfies the dependence on $\Omega^2$. However, as stated in that work, their study of imaginary angular velocity is based on the series expansion $\ln (1-z)=-\sum_{n=1}^{\infty}z^n/n$ for $|z|\leq 1$. Performing the continuation $\Omega^2=-\omega^2$ directly on this basis effectively uses the series expansion in the region $|z|>1$, which is entirely incorrect. However, direct calculation presents other problems, as introducing a real angular velocity is equivalent to introducing an effective chemical potential. For a bosonic system, the chemical potential must satisfy $|\mu|\leq m$. Since the $W^\pm$ gauge bosons in our calculation are massless, this leads to an infrared singularity. This singularity originates from a violation of causality because, from the conventional view $k_n=2\pi n/L$, low-momentum modes imply the system's boundary is at infinity, which contradicts the finiteness of a rotating system. There are two approaches to address this singularity \cite{Fujimoto:2021xix}. One is to still consider transverse momentum $k_\perp$ as continuously varying and deem modes with $k_\perp<\omega$ as unphysical, applying an infrared cutoff to this part. The other is to impose boundary conditions so that the system's lowest transverse momentum $k_\perp^\text{min}$ self-consistently exceeds $\omega$.

Since it is difficult to establish a direct relationship between low-momentum modes and superluminal behavior (i.e., determining at which momentum the mode becomes acausal) and one can only simplistically assume that the singular transverse momentum states with $k_\perp<\omega$ are superluminal, while still treating transverse momentum $k_\perp$ as continuous (which inherently assumes an infinite boundary), this cutoff method in our calculation would directly lose the crucial contribution from the LLL, which is clearly unphysical. Therefore, we adopt the second approach by imposing the following boundary condition:
\begin{equation}
 F(-\lambda_k^l,|l|+1,X)=0,\quad X=\frac{1}{2}gH R^2
\end{equation}
to obtain the finite-temperature part of the potential at $r=0$ (details in Appendix~\ref{B}):
    \begin{align}
         V_T&(\vec{0})=\frac{T}{(2\pi)^2}\sum_{s=\pm}\sum_k \frac{gH}{ N_k^{0}}\int_{-\infty}^{+\infty}\frac{dk_z}{2\pi}\notag\\
         &\ln \left(1-e^{-\beta(\omega_s^0+s\omega)+i\phi}\right)+\ln \left(1-e^{-\beta(\omega_s^0-s\omega)-i\phi}\right).
    \label{thermo}
    \end{align}
    
\begin{figure}[htbp]
    \centering
    \includegraphics[width=8.6cm]{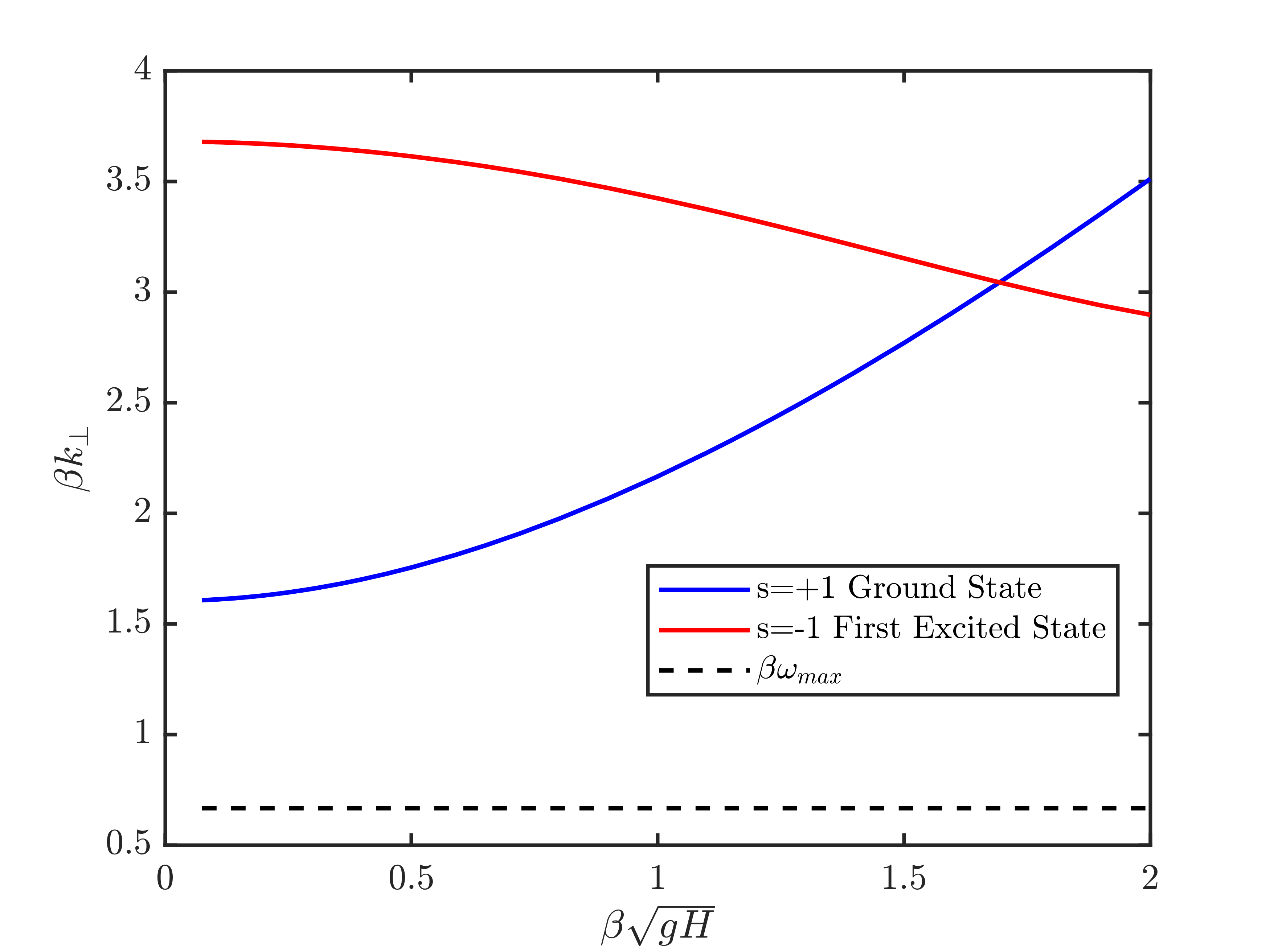}
    \caption{The normalized transverse momenta $\beta k_\perp$ for the first excited state ($s=-1$) and the ground state ($s=1$) as functions of $\beta\sqrt{gH}$, compared with $\beta\omega_{\text{max}}$.}
    \label{fig8}
\end{figure}

Here,
   \begin{gather}
   N_k^0=\frac{1}{2\pi}\int_0^Xdx F(-\lambda_k^0,1,x)^2e^{-x},\\
   \omega_s^0=\sqrt{k_z^2+(2\lambda_k^0+2s+1)gH}.
\end{gather}
In subsequent calculations, we set $TR=1.5$ to be consistent with the parameters used in reference \cite{Chen:2022smf}. It can be verified that in the limit $H\rightarrow0$, the result reduces to that in \cite{Chen:2022smf} (see Appendix~\ref{C}). Now we need to discuss whether imposing boundary conditions self-consistently resolves the infrared divergence issue. Firstly, for the Savvidy model, its LLL inherently possesses a long-wavelength instability in the quadratic approximation, so theoretically, imposing boundary conditions cannot resolve the intrinsic instability of the Savvidy model itself. We need to verify whether, for higher Landau levels, the transverse momenta $k_\perp$ for the first excited state with $s=-1$ and the ground state with $s=1$, which are $\sqrt{(2\lambda_2^0-1)gH}$ and $\sqrt{(2\lambda_1^0+1)gH}$ respectively, are greater than $\omega_\text{max}=1/R$. As shown in Fig.~\ref{fig8}, within the range of $\beta\sqrt{gH}$ values achievable in subsequent calculations, the boundary condition indeed resolves this problem.

However, two issues remain in our system. First, regarding the long-wavelength instability of the LLL, we follow the convention of previous literature and take the real part of this contribution. Second, there is the singularity arising from the complex chemical potential. When both the real angular velocity and the Polyakov loop are non-zero, the modification to the Matsubara frequencies becomes a complex number with non-zero real and imaginary parts: $-\beta s \omega+i\phi$, equivalent to introducing a complex chemical potential. Consequently, the contributions from the two polarization modes $s=\pm$ to the potential are not real. In the absence of a chromomagnetic background, since the single-particle spectrum $|\vec{k}|$ has no spin dependence, these contributions cancel overall. However, due to the coupling between the chromomagnetic field and spin, $\omega_s^0$ now differs with spin, leading to an imaginary part in the overall potential. Here, we still study the real part of the potential. Nevertheless, subsequent calculations of the Polyakov loop potential reveal that for any angular velocity, Although a singularity emerges in the local vicinity, the physical minimum itself remains well-defined and is located at $\phi=0$ and $2\pi$. 

We have performed the numerical integration and summation of Eq. (\ref{thermo}) for real $\omega$. The sums over $k$ and the $k_z$ integration are cut off at sufficiently large numbers, and convergence is confirmed. It should be noted that numerically achieving $H=0$ remains unfeasible. We set a lower limit of $\beta\sqrt{gH}=0.075$, and the numerical results for the potential at this value show convergence to the result without chromomagnetism.
\begin{figure*}[t]
    \centering
    % 第一行：图片 1 和图片 2
    \begin{subfigure}[b]{0.45\textwidth}
        \includegraphics[width=\textwidth]{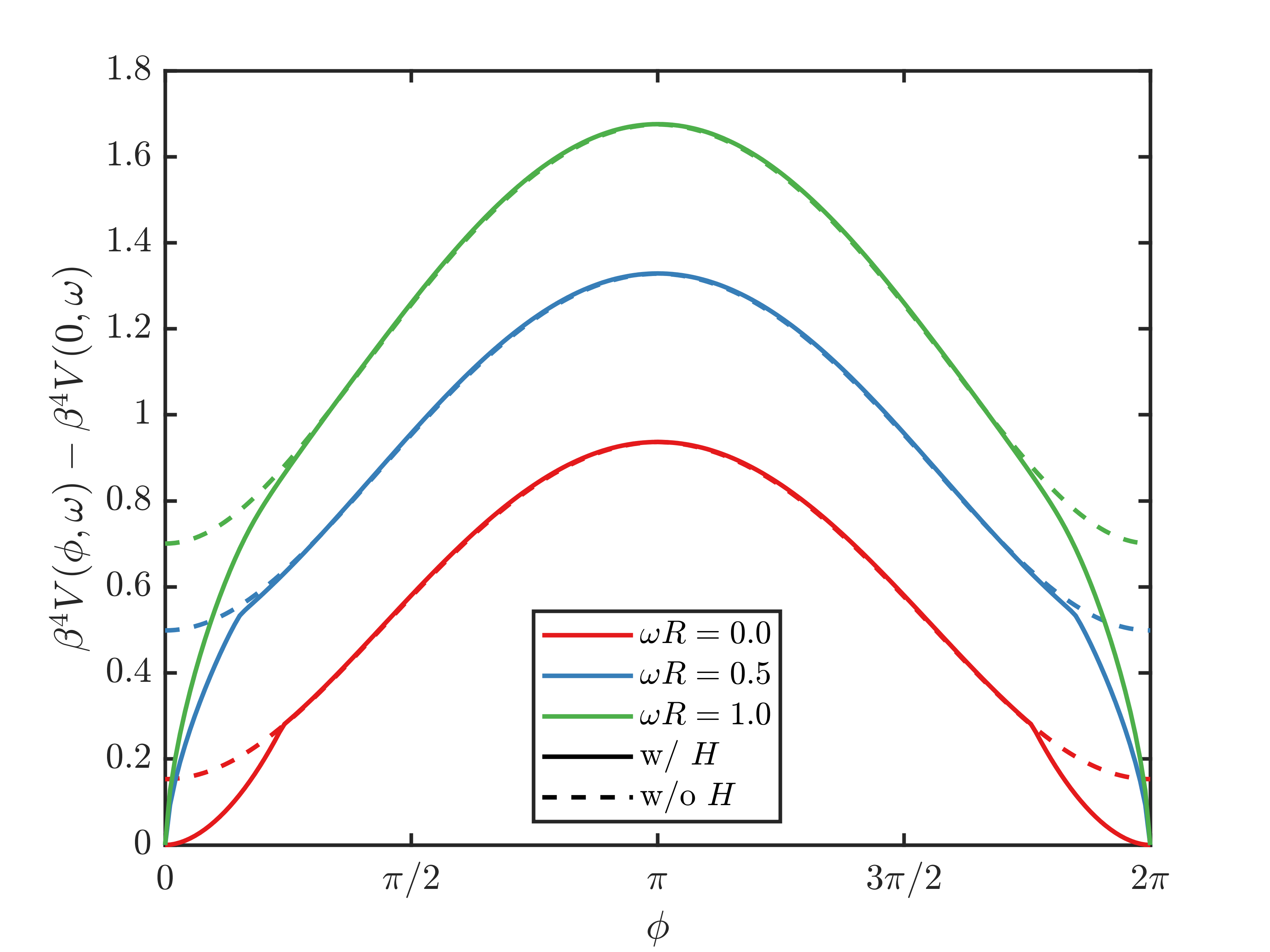}
        \phantomcaption
    \end{subfigure}
    \hspace{-0.5cm} % 负间距缩小水平间隔
    \begin{subfigure}[b]{0.45\textwidth}
        \includegraphics[width=\textwidth]{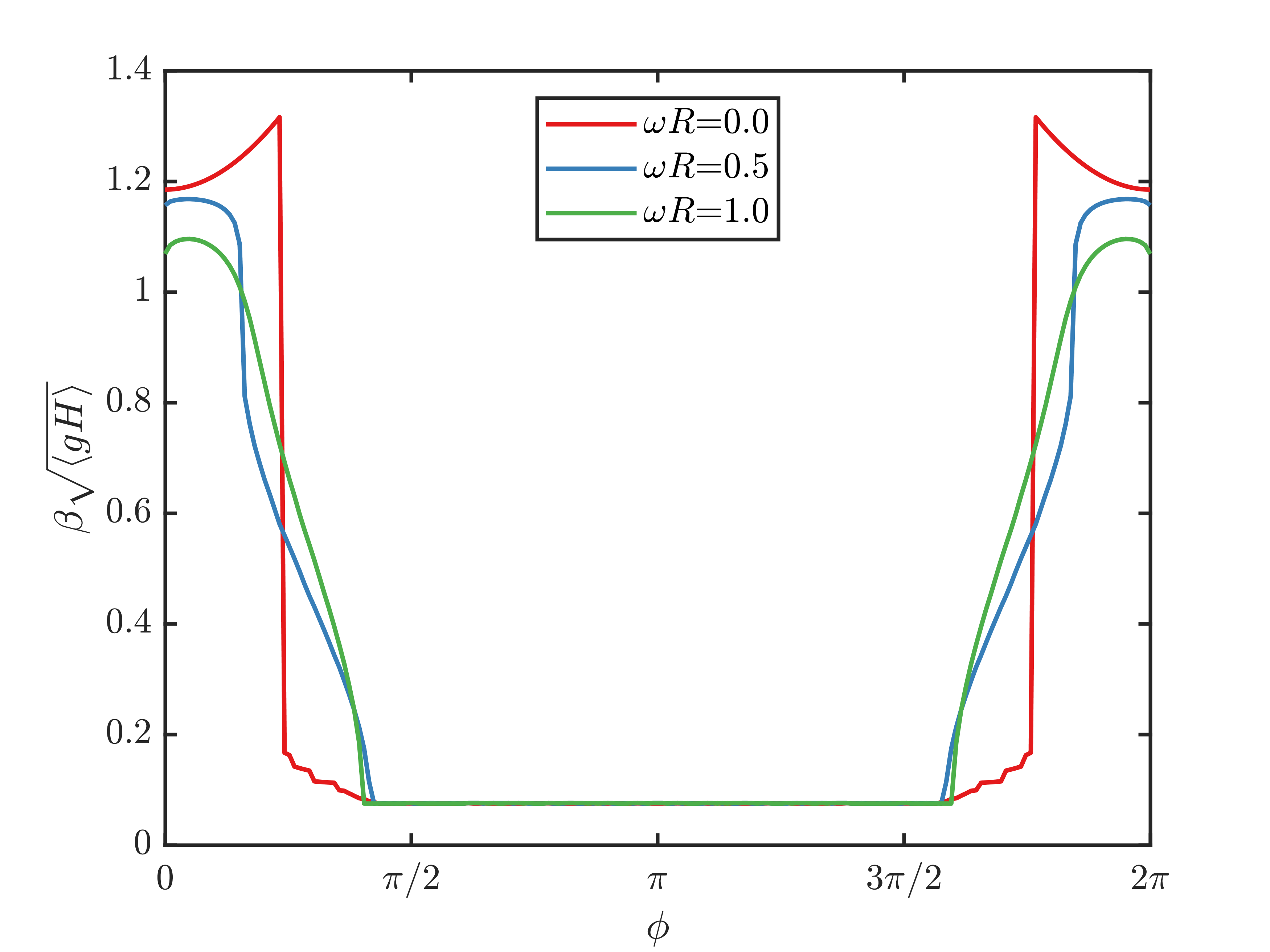}
        \phantomcaption
    \end{subfigure}
 % \vspace*{-0.5cm}
 \caption{Evolution of the dimensionless Polyakov loop potential $\beta^4 V$ (left) and the corresponding scaled variable $\beta \sqrt{\langle gH \rangle}$ (right) with $\phi$ for several values of the real angular velocity $\omega$ at $T=10\Lambda$ and $r=0$.}
 \label{fig9}
\end{figure*}

Fig.~\ref{fig9} shows the evolution of the Polyakov loop potential (solid lines with chromomagnetic field, dashed lines without) with increasing $\omega$ at $r=0$ and $T=10\Lambda$, along with the corresponding scaled variable $\beta \sqrt{\langle gH \rangle}$. The potential curve with a chromomagnetic background remains lower than that without. In regions where the two curves nearly coincide, $\beta \sqrt{\langle gH \rangle}$ is near our set lower limit and can be considered as vanished. The potential minima are located at $\phi=0$ mod $2\pi$ for any $\omega$, indicating the system remains in the deconfined phase. However, a point of caution: when $\omega\neq0$, the minimum of our Polyakov loop potential is a cusp (unlike the case with imaginary angular velocity), where the first derivative does not exist. This implies the potential curvature around the minimum, which defines the Debye screening mass, lacks a traditional definition. The reason is that the local thermodynamic potential at the minimum possesses an imaginary part due to the complex chemical potential. In contrast, for $\omega=0$, the effective chemical potential is always imaginary, avoiding the singularity from a real effective chemical potential. Although long-wavelength instability ($k_\perp^\text{LLL}<0$) persists in the LLL near $\phi=0$, the minimum is a stable point with zero first derivative. Based on the preceding discussion, the emergence of a cusp in the Polyakov loop potential under real rotation manifests in a fundamental alteration of local stability and suggests a novel singularity within the phase structure. This behavior indicates that the system's properties near the minimum may not be fully captured by conventional equilibrium thermodynamics, as the effective complex chemical potential arising from the coupling between rotation, spin, and the chromomagnetic field introduces an imaginary component of Polyakov loop potential. 

\begin{figure}[htbp]
    \centering
    \includegraphics[width=8.6cm]{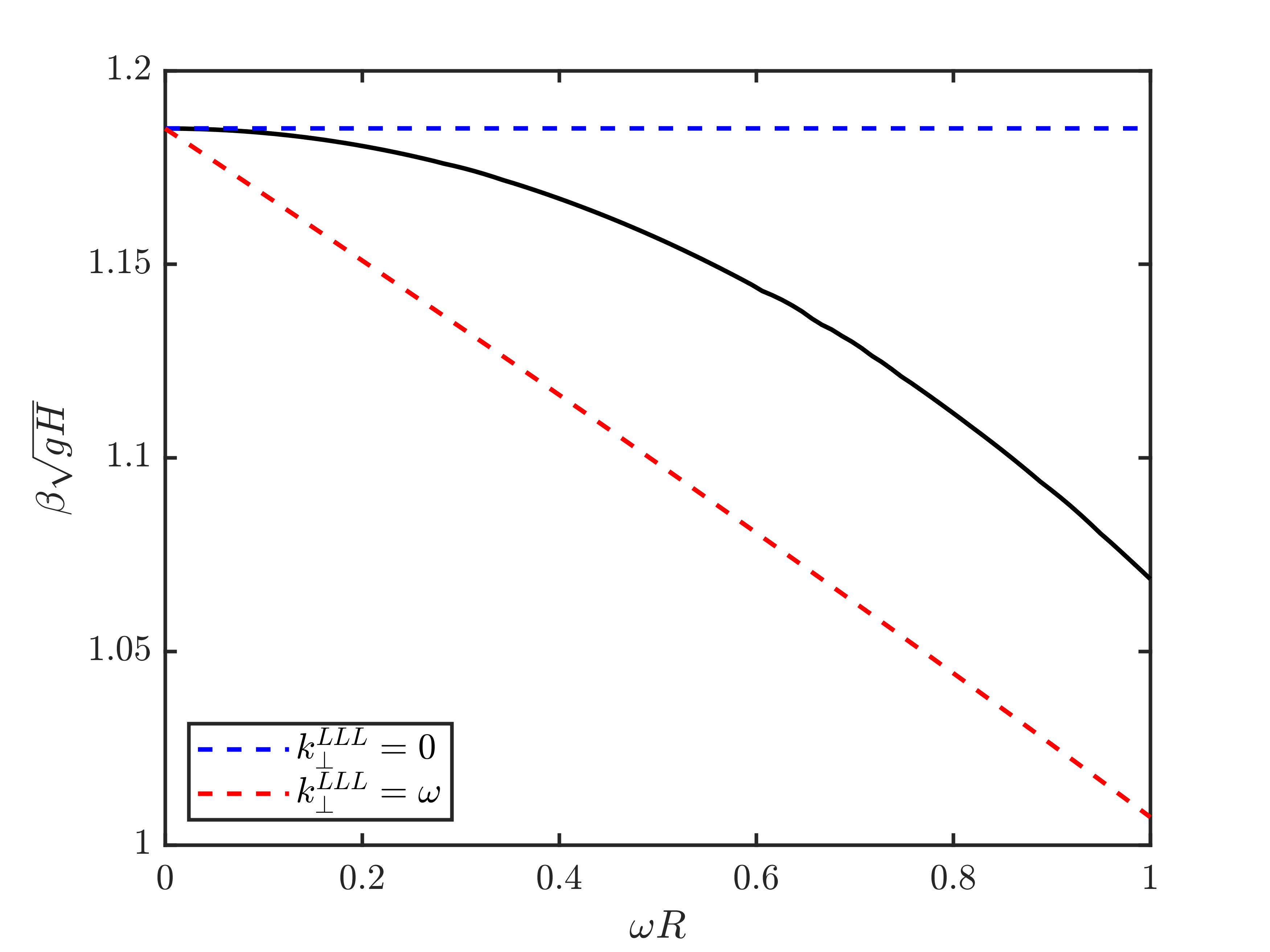}
    \caption{Evolution of the scaled chromomagnetic condensate $\beta\sqrt{gH}$ as a function of the real angular velocity $\omega$, with dashed lines indicating $k_{\perp}^\text{LLL}=0$ and $k_{\perp}^\text{LLL}=\omega$.}
    \label{fig10}
\end{figure}

Fig.~\ref{fig10} shows the variation of the chromomagnetic condensate with real angular velocity (black solid line). In contrast to the previous case with imaginary angular velocity, for $T=10\Lambda$, the chromomagnetic condensate exists across the entire $\omega$ range. This is a consequence of the boundary conditions, which change the Landau level quantum numbers from uniformly spaced $m \in \mathbb{N}$ to non-uniformly spaced $\lambda_k$, increasing the energy gaps and significantly suppressing contributions from higher Landau levels. Consequently, the system is dominated by the LLL. As the angular velocity increases, the chromomagnetic condensate becomes smaller. In previous studies \cite{Cea:2007yv}, the coupling of the chromomagnetic field to the quark system alters the chiral condensate and the chiral phase transition temperature, suggesting a possibility that rotation-induced changes in the chromomagnetic field could indirectly affect the chiral phase transition. The blue dashed line represents the chromomagnetic condensate value when $k_{\perp}^\text{LLL}=0$, and the red dashed line represents the value when $k_{\perp}^\text{LLL}=\omega$. Our calculated chromomagnetic condensate lies below the black dashed line and above the red dashed line, indicating $k_{\perp}^\text{LLL}\in [0,\omega)$. Under this treatment, the long-wavelength instability persists, but its nature changes from the traditional type caused by $k_{\perp}^\text{LLL}<0$ leading to an imaginary spectrum, to a type caused by $k_{\perp}^\text{LLL}<\omega$ leading to a negative energy spectrum.

It should be emphasized that the system considered in this work does not suffer from thermodynamic instability caused by a radial drift force~\cite{Fukushima:2024tkz}. The reason is that, for the same physical background, although the form of the gauge potential is different in different reference frames, the corresponding force-balance relation is the same.

In the \emph{local inertial frame}, we choose the gauge potential
\begin{equation}
    \bar{A}_a=\left(0,\frac{H}{2}y,-\frac{H}{2}x,0\right),
\end{equation}
which corresponds to
\begin{equation}
    \mathbf{\bar B}=(0,0,-H),\qquad
    \mathbf{\bar E}=0.
\end{equation}
Since the system is at rest in the local inertial frame, \(\mathbf v=0\), the Lorentz force
\begin{equation}
    \mathbf F=q(\mathbf{\bar E}+\mathbf v\times\mathbf{\bar B})=0.
\end{equation}
Therefore, there is no radial drift force in this frame.

In the \emph{curved frame}, the corresponding gauge potential is
\begin{equation}
    A_\mu
    =e_\mu^{\,a}\bar A_a
    =\left(-\frac{H\omega r^2}{2},\frac{H}{2}y,-\frac{H}{2}x,0\right),
    \  r^2=x^2+y^2.
\end{equation}
From this we obtain
\begin{equation}
    \mathbf B=(0,0,-H),\quad
    \mathbf E=(H\omega x,H\omega y,0).
\end{equation}
On the other hand, the local velocity field in the rotating frame is
\begin{equation}
    \mathbf v=\boldsymbol{\omega}\times\mathbf r=(-\omega y,\omega x,0),
\end{equation}
and therefore
\begin{equation}
    \mathbf F=q(\mathbf E+\mathbf v\times\mathbf B)=0.
\end{equation}
This shows that, although \(\mathbf v\neq 0\) in the curved frame, the electric field induced by the time component \(A_0\) exactly cancels the radial drift term produced by the magnetic field. Therefore, there is no unbalanced radial drift force in the system, and the thermodynamic instability we considered here comes from Savvidy vaccum under real rotation.

\section{summary and outlook}
\label{sec:6}
The reappearance of non-perturbative physics in the perturbative regime driven by an imaginary angular velocity is highly intriguing. As our results demonstrate, the imaginary angular velocity not only induces a perturbative confinement phase but also triggers chromomagnetic condensation, exerting significant and non-trivial influences on the phase diagram of the perturbative confinement transition.

Through analysis, we decompose the effective potential into two components: the chromagnetic-favoring potential $V_H$ and the chromagnetic-suppressing potential $V_{\text{nonH}}$. We systematically study the effects of temperature and imaginary angular velocity on these components, thereby elucidating the microscopic mechanism by which $\tilde{\Omega}$ induces the chromomagnetic condensate: the imaginary angular velocity shifts the minimum of $V_H$ along the $\phi$ direction while significantly weakening the suppressing effect of $V_{\text{nonH}}$ on chromomagnetism, thereby altering the competitive balance between them. This leads us to obtain a first-order phase boundary with explicit temperature dependence, identified by the discontinuous jump of the order parameters $|L|$ and $H$ at the critical point. This phase boundary asymptotically approaches the theoretical value for the case without a chromomagnetic background, $\tilde{\Omega}_c = \pi/\sqrt{3}$, at high temperatures.

We thereby construct a new phase diagram, identifying and confirming that within the $\tilde{\Omega}$-$T$ phase structure, a more extensive region at high temperatures belongs to the deconfined phase. The perturbative confinement transition consequently exhibits a richer phase structure, which directly manifests the complex coupling between the chromomagnetic field and the Polyakov loop. Under the reasonable assumption of a continuous phase boundary, by analyzing its trend before entering the non-perturbative temperature regime, we speculate that non-monotonic behavior of the phase boundary may occur in the non-perturbative regime, revealing the possible complex non-perturbative effects therein. This could significantly impact the adiabatic continuity evolution path connecting the perturbatively confined phase and the conventional confined phase. Given the second-order nature of the non-perturbative deconfinement transition in $\mathrm{SU}(2)$ theory, a critical end point (CEP) may exist where it meets the first-order perturbative confinement transition. 

Finally, we investigate the effects of a real angular velocity $\omega$. Adopting the conventional treatment, we impose boundary conditions to preserve causality and take the real part for the Savvidy instability (LLL) in the quadratic approximation. Our study finds that the effective complex chemical potential introduced by the real angular velocity, due to the coupling between the chromomagnetic field and spin, prevents the imaginary contributions from the two spin polarization states ($s=\pm$) from canceling each other, resulting in a non-zero imaginary part in the total potential. This leads to a \textbf{cusp} (discontinuous first derivative) at the minimum of the Polyakov loop potential, implying the traditional Debye screening mass cannot be defined at this point. This phenomenon reflects the possible complex non-equilibrium evolution when the chromomagnetic field, rotation, and spin are coupled. Furthermore, we observe that the strength of the chromomagnetic condensate decreases with increasing real angular velocity. Throughout this process, the long-wavelength instability of the LLL in the quadratic approximation persists, but its character changes: the instability threshold shifts from the traditional condition $k_{\perp}^\text{LLL} < 0$ (leading to an imaginary spectrum) to $k_{\perp}^\text{LLL} < \omega$ (leading to a negative energy spectrum).

 Imaginary angular velocity can open a new pathway to study non-perturbative physics within a perturbative framework. A natural extension is to introduce dynamical quarks to explore whether imaginary rotation can also induce chiral symmetry breaking in the perturbative regime. If realized, it would suggest that key non-perturbative features—including confinement, chiral symmetry breaking, and chromomagnetic condensation—could be systematically accessed via perturbative methods under imaginary rotation. Furthermore, generalizing the analysis to $\mathrm{SU}(3)$ gauge theory would provide a more direct link to QCD. Finally, the predictions from our $\tilde{\Omega}$-$T$ phase diagram, particularly the induced chromomagnetic condensate and the first-order transition boundary, offer concrete targets for future lattice simulations, which are crucial for mapping the complete phase structure of rotating gauge theories.

\begin{acknowledgments}
We acknowledge helpful discussions with Hao-Lei Chen, Defu Hou and Xuguang Huang. This work is supported in part by the National Natural Science Foundation of China (NSFC) Grant Nos: 12235016, 12221005, 12505148. Recently, we learned that Ref. \cite{Chen:2026ced} addresses a similar topic, which was posted on arXiv on the same day. 
\end{acknowledgments}

\appendix
\section{\MakeUppercase{the rotation-modified Lagrangian density}}
\label{A}
Since the derivative operator appears in the Lagrangian only in the form $\partial_{[\mu}\square_{\nu]}$, the transformation of the entire system satisfies the identity (\ref{rotaM}). After changing the ordinary derivatives in $B_{\mu\nu}$ to covariant derivatives and introducing the Polyakov background field, we have:
\begin{equation}
    B_{\mu\nu}\rightarrow\partial_{[\mu} B_{\nu]}+\delta_{[\mu 0}\  R_0 B_{\nu]}.
\end{equation}
Since
\begin{align}
    R_0B_k=&-\omega\epsilon_{ji3}x_j\partial_iB_k+\omega\epsilon_{k3i}B_i\notag\\
    =&\frac{1}{2}H\left(-\omega\epsilon_{ji3}x_j\partial_i(\epsilon_{kj3}x_j)+\omega\epsilon_{k3i}\epsilon_{ij3}x_j\right)\notag\\
    =&\frac{1}{2}H\left(-\omega\epsilon_{ji3}\epsilon_{ki3}x_j+\omega\epsilon_{k3i}\epsilon_{ij3}x_j\right)=0,
\end{align}
the tensor $B_{\mu\nu}$ remains unchanged, and consequently the term $B_{\mu\nu}B^{\mu\nu}$ that gives rise to $H^2$ is also unchanged. Ultimately, the effect of rotation originates primarily from the kinetic term $\hat{Q}^+_{\mu\nu}\hat{Q}^{-,\mu\nu}$. With
\begin{equation}
    D_\mu\rightarrow D_\mu+\delta_{\mu 0}R_0, 
\end{equation}
we compute the additional part $\mathcal{L}_{\text{add}}$ of the Lagrangian to reduce the computational burden, i.e., $\mathcal{L}_0\rightarrow \mathcal{L}_0+\mathcal{L}_{\text{add}}$, where
\begin{gather}
    \hat{Q}^+_{\mu\nu} \rightarrow \hat{Q}^+_{\mu\nu}+\delta_{0[\mu} R_0Q^+_{\nu]}, \\
    \hat{Q}^-_{\mu\nu} \rightarrow \hat{Q}^-_{\mu\nu}+\delta_{0[\mu}R_0Q^-_{\nu]}.
\end{gather}
Thus, we obtain
\begin{equation}
    -\frac{1}{2}\hat{Q}^+_{\mu\nu}\hat{Q}^{\mu\nu,-}\rightarrow -\frac{1}{2}\hat{Q}^+_{\mu\nu}\hat{Q}^{\mu\nu,-}+\mathcal{L}_{\text{add}},
\end{equation}
with
\begin{equation}
    \mathcal{L}_{\text{add}}=-\left[\hat{Q}^+_{0i}R_0Q^{i,-}+\hat{Q}^{0i,-}R_0 Q^{+}_i +(R_0Q^{+}_i)(R_0Q^{i,-})\right].
\end{equation}
and noting the symmetry $\mathcal{L}_{\text{add}}=\mathcal{L}_{\text{add}}^\dagger$, the equations of motion (EOM) for $Q^+$ and $Q^-$ are conjugate to each other. By solving the Euler-Lagrange equation
\begin{equation}
    \left[\frac{\partial}{\partial Q^-_\nu}-\partial_\mu\frac{\partial}{\partial(\partial_\mu Q^-_\nu)}\right]\mathcal{L}_{\text{add}},
\end{equation}
we obtain the additional terms that need to be added to the EOM without rotation.
\begin{widetext}
For the temporal component $\nu=0$:
\begin{equation}
     -i\omega\hat{L}_z(\partial_iQ_i^+)+igB_iR_0Q_i^+=-i\omega\hat{L}_z(D_iQ_i^+).
\end{equation}

For the spatial components $\nu=i$:
\begin{equation}
    (2i\omega\hat{L}_zD_0+\omega^2\hat{L}_z^2)Q_i^+ +2(D_0-i\omega\hat{L}_z)\omega\epsilon_{ij3}Q_j^++\omega^2(\delta_{ik}-\delta_{i3}\delta_{k3})Q_k^++\omega(-i\hat{L}_z\partial_i-\epsilon_{ij3}\partial_j-igB_i i\hat{L}_z)Q_0^+
\end{equation}
It can be seen that the inclusion of angular velocity significantly affects the equations of motion, particularly through a strong coupling with $Q_0$. If, under a suitable gauge condition, the coupling between $Q_0$ and the spatial components can be eliminated, it would greatly simplify our study of the spin gluons $Q_\pm$. Substituting the additional terms into the original EOM: 
\begin{equation}
    D_\mu D^\mu Q^{\nu,+}-D^\nu D_\mu Q^{\mu,+}+2igQ_\lambda^+B^{\nu\lambda}=0
\end{equation}
and analyzing the temporal component $\nu=0$:
\begin{align}
    &D_\mu D^\mu Q^{0,+}-D^0 D_\mu Q^{\mu,+}-i\omega\hat{L}_z(D_iQ_i^+)\notag\\
    =&D_i D^i Q^{0,+}-D^0 D_i Q^{i,+}-i\omega\hat{L}_z(D_iQ_i^+)\notag\\
    =&D_i D^i Q^{0,+}+(D_0-i\omega\hat{L}_z)^2 Q^{0,+}-(D_0-i\omega\hat{L}_z)^2 Q^{0,+}-(D_0-i\omega\hat{L}_z)(D_iQ_i^+)\notag\\
    =&(D_\mu-\delta_{0\mu}i\omega\hat{L}_z) (D^\mu-\delta_{0\mu}i\omega\hat{L}_z)Q^{0,+}-(D_0-i\omega\hat{L}_z)(D_\mu-\delta_{0\mu}i\omega\hat{L}_z) Q^{\mu,+}=0.
\end{align}
Analyzing the part involving $Q_0^+$ in the spatial components:
\begin{align}
    -D^i(D_\mu Q^{\mu,+})+\omega(-i\hat{L}_z\partial_i-\epsilon_{ij3}\partial_j-igB_i i\hat{L}_z)Q_0^+=-D^i(D_\mu Q^{\mu,+}-i\omega \hat{L}_z Q_0^+).
\end{align}
By replacing the covariant derivative as $D_0\rightarrow D_0=\partial_0+igB_0-i\omega\hat{L}_z$ and choosing the gauge condition:
\begin{equation}
    D_\mu Q^{\mu,+}=0,
\end{equation}
we can eliminate the coupling between $Q_0$ and $Q_i$ in the equation with the gauge-fixing term in this form:
\begin{equation}
    \mathcal{L}_\text{gf}(B,Q) = -\frac{1}{2}(\partial_\mu Q^{\mu,3}- \omega i\hat{L}_zQ^{0,3})^2
    -(D_\mu Q^{\mu,+})(D_\mu^* Q^{\mu,-}),.
\end{equation}
Thus, we have successfully decoupled the temporal and spatial components in the equations of motion. In this new gauge, the additional terms in the equations of motion become:
\begin{align}
    \nu=1:& \ \left[2i\omega\hat{L}_zD_0+\omega^2\hat{L}_z^2+\omega^2\right]Q_1^++2(D_0-i\omega\hat{L}_z)\omega Q_2^+,\\
    \nu=2:& \ \left[2i\omega\hat{L}_zD_0+\omega^2\hat{L}_z^2+\omega^2\right]Q_2^+-2(D_0-i\omega\hat{L}_z)\omega Q_1^+,\\
    \nu=3:& \ \left[2i\omega\hat{L}_zD_0+\omega^2\hat{L}_z^2\right]Q_3^+.
\end{align}
Transforming the additional terms in the EOM to the spin components $Q_\pm^+=Q_1^+\pm iQ_2^+$, the equations take the form:
\begin{equation}
    \begin{pmatrix}
    \hat{L}_1 & \hat{L}_2\\
    -\hat{L}_2 & \hat{L}_1
    \end{pmatrix}
    \begin{pmatrix}
        Q_1^+\\Q_2^+
    \end{pmatrix}=X^{-1}
    \begin{pmatrix}
    \hat{L}_1 & \hat{L}_2\\
    -\hat{L}_2 & \hat{L}_1
    \end{pmatrix}
    XX^{-1}
    \begin{pmatrix}
        Q_1^+\\Q_2^+
    \end{pmatrix}=
    \operatorname{diag}(\hat{L}_1 -i \hat{L}_2,\hat{L}_1+i\hat{L}_2 )
    \begin{pmatrix}
        Q_+^+\\Q_-^+
    \end{pmatrix}=0,
\end{equation}
where
\begin{gather}
    X=\begin{pmatrix}
        1&1\\
        -i&i
    \end{pmatrix},\quad X^{-1}=
    \frac{1}{2}\begin{pmatrix}
        1&i\\
        1&-i
    \end{pmatrix},\\
    \hat{L}_1 = 2i\omega\hat{L}_zD_0+\omega^2\hat{L}_z^2+\omega^2,\quad \hat{L}_2=2(D_0-i\omega\hat{L}_z)\omega.
\end{gather}
These additional terms will shift the time derivative $D_0$ in the original equation. For $\hat{L}_1 -s i \hat{L}_2$ with $s=\pm  1$, we have:
\begin{align}
    &(iD_0)^2+\hat{L}_1 -is\hat{L}_2\notag\\
    =&(iD_0)^2+2i\omega\hat{L}_zD_0+\omega^2\hat{L}_z^2+\omega^2-2is(D_0-i\omega\hat{L}_z)\omega\notag\\
    =&(iD_0+\omega\hat{L}_z)^2+\omega^2-2s(iD_0+\omega\hat{L}_z)\omega\notag\\
    =&\left[iD_0+\omega(\hat{L}_z-s) \right]^2.
\end{align}
\end{widetext}
Thus, the angular velocity shifts the energy levels according to Eq. (\ref{wn}). The eigenmodes remain unchanged, and the coupling forms between the Landau level quantum number and spin, and between the angular momentum quantum number and spin, are $m+s$ and $l-s$, respectively. Substituting into the gauge condition, this exactly corresponds to the recurrence relation between $L_m^{l}$ and $L_{m+s}^{l-s}$, indicating that the chosen orientation of the magnetic field is appropriate.

In the imaginary-time formalism of finite temperature field theory, the zero component of all covariant vectors, such as the time component of four-dimensional spacetime, is transformed with $t=-i\tau$. Similarly, the zero component of the electromagnetic four-vector potential becomes $B_0\rightarrow-iB_0=-i(g\beta)^{-1}\phi$. Taking the imaginary angular velocity $\omega=i\Omega$, we obtain:
\begin{equation}
    D_0\rightarrow i\partial_\tau+\phi/\beta+\Omega\left(\hat{L}_z-s\right).
\end{equation}
\section{\MakeUppercase{Gauge invariance}}
\label{D}
As pointed out in Ref.~\cite{Fukushima:2020ncb, Fukushima:2024tkz}, for a Dirac field in a rotating frame coupled to an external magnetic field, the gauge symmetry is no longer manifest in the form
\begin{equation}
    \left(i\slashed{D}+\gamma^0\vec{\Omega}\cdot\vec{J}-m\right)\psi=0.
\end{equation}
Indeed, the angular momentum operator $\vec{J}$ is not gauge invariant. However, for the SU(2) Yang--Mills theory,
\begin{equation}
\begin{gathered}
A_\mu = A_\mu^a \frac{\sigma^a}{2}, \\
F_{\mu \nu}^a = \partial_{[\mu} A_{\nu]}^a + g \epsilon^{abc} A_{\mu}^b A_{\nu}^c, \\
F_{\mu \nu} = F_{\mu \nu}^a \frac{\sigma^a}{2} = \partial_{[\mu} A_{\nu]} - i g A_{[\mu} A_{\nu]}, \\
\mathcal{L} = -\frac{1}{4} \text{Tr}(F_{\mu \nu} F^{\mu \nu}).
\end{gathered}
\end{equation}
In a rotating background, the theory can be described either in the \emph{curved frame}, or through the vierbein formalism in the \emph{local rest/inertial frame}. Denote by \(A,D\) the field and derivative in the curved frame, and by \(\bar A,\bar D\) the corresponding quantities in the local frame. Then in the curved frame,
\begin{equation}
    S[A]=\int d^4x\, g^{\mu\nu}g^{\alpha\beta}\text{Tr}(F_{\mu\alpha}F_{\nu\beta}),
\end{equation}
where
\begin{equation}
    F_{\mu \nu}=D_{[\mu}A_{\nu]}-igA_{[\mu}A_{\nu]}
    =\partial_{[\mu}A_{\nu]}-igA_{[\mu}A_{\nu]},
\end{equation}
for \(D_\mu A_\nu=\partial_\mu A_\nu+G_{\mu \nu}^\rho A_\rho\), and \(G_{\mu \nu}^\rho\) is the usual Christoffel connection in curved space-time
\begin{equation}
    G_{\mu\nu}^{\rho}=G_{\nu\mu}^{\rho},
\end{equation}
which automatically vanishes in the antisymmetric combination.
In the local rest/inertial frame,
\begin{equation}
    S[\bar A]=\int d^4x\, \eta^{ac}\eta^{bd}\text{Tr}(\bar F_{ab}\bar F_{cd}),
\end{equation}
where
\begin{equation}
    \begin{gathered}
        \bar{A}_a=e_a^\mu A_\mu, \\ 
    \bar{F}_{ab}=e_a^\mu e_b^\nu F_{\mu\nu}=\bar{D}_{[a}\bar{A}_{b]}-ig\bar{A}_{[a}\bar{A}_{b]},\\ 
    \bar{D}_a\bar{A}_b=e_a^\mu D_\mu \bar{A}_b=e_a^\mu(\partial_\mu \bar{A}_b+\Gamma_{\mu b c}\bar{A}^c).
    \end{gathered}
\end{equation}
Here \(e_a^\mu\) is called the tetrad, and \(\Gamma_{\mu b c}\) is the spin connection. Using \(\eta _{ab}=e_a^\mu e_b^\nu g_{\mu\nu}\), the relation between the expressions in different reference frames is
\begin{equation}
\mathcal{L}=g^{\mu\nu}g^{\alpha\beta}F^k_{\mu\alpha}F^k_{\nu\beta}=\eta^{ac}\eta^{bd}\bar{F}^k_{ab}\bar{F}^k_{cd}.
\end{equation}
Therefore \(S[A]=S[\bar{A}]\). In the curved frame, we regard the gauge transformation as a transformation in the \textbf{internal SU(2) space}, which itself does not change because of the change of the space-time reference frame. The form of the gauge transformation is still
\begin{equation}
    \begin{gathered}
        A_{\mu}^U=U\left(A_{\mu}+\frac{i}{g}\partial_{\mu}\right)U^\dagger, \\
    F_{\mu \nu}^U=UF_{\mu \nu}U^\dagger.
    \end{gathered}
\end{equation}
Since
\begin{equation}
    \text{Tr}(F_{\mu \nu}^U F_{\rho \sigma}^U)=\text{Tr}(F_{\mu \nu}F_{\rho \sigma}),
\end{equation}
the action is gauge invariant in the curved frame. Therefore, the local frame must also preserve gauge invariance. Using the vierbein to project it to the local inertial basis \(\bar A_a\), because this local basis itself is rotating and depends on the coordinates, the explicit gauge transformation of \(\bar{A}\) will carry extra structure. The gauge transformation law is no longer the simple flat-space form. For rotation around the \(z\) axis, take
\begin{equation}
    e_a^\mu=\delta_a^\mu+\omega \delta_a^0 \epsilon_{0\mu\nu 3}x_\nu,
\end{equation}
then
\begin{equation}
    \bar A_a^U
    =e_a^\mu A_\mu^U
    =U\left(\bar A_a+\frac{i}{g}\partial_a+\delta_a^0\frac{\omega}{g}\hat L_z\right)U^\dagger.
\end{equation}
That is to say, in the local frame, \(\hat L_z\) is precisely a part of the gauge transformation law, so it is not an arbitrarily inserted breaking term, but a term required for maintaining the gauge-covariant structure in this reference frame. It satisfies gauge invariance
\begin{equation}
    S[\bar{A}^U]=S[A^U]=S[A]=S[\bar{A}].
\end{equation}
Furthermore, under the background-field decomposition
\begin{equation}
    \bar A_a=B_a+Q_a,
\end{equation}
the background-field gauge transformation is
\begin{equation}
    \begin{gathered}
        B_\mu^U
    =U\left(B_\mu+\frac{i}{g}\partial_\mu+\delta_\mu^0\frac{\omega}{g}\hat L_z\right)U^\dagger, \\
    Q_\mu^U=UQ_\mu U^\dagger.
    \end{gathered}
    \label{GT1}
\end{equation}
Since the transformation of \(Q\) is homogeneous, the expansion of the action in powers of \(Q\),
\begin{equation}
    S[B+Q]=S[B]+S_2[B,Q]+\cdots
\end{equation}
satisfies background gauge covariance order by order:
\begin{equation}
    S_n[B,Q]=S_n[B^U,Q^U].
\end{equation}
This is precisely the fundamental reason why the background-field method guarantees the gauge invariance of the one-loop effective action.
In our article, we perform the one-loop calculation for the quadratic fluctuation of \(Q\), which is obviously gauge invariant.
\begin{equation}
    \mathcal{L}_\text{1-loop}(B,Q)=\mathcal{L}_\text{1-loop}(B^U,Q^U).
\end{equation}
We have the redundancy of the quantum gauge symmetry \(\bar{U}\):
\begin{equation}
    \begin{gathered}
         B_\mu^{\bar{U}}= B_\mu,\\
   Q_{\mu}^{\bar{U}} =U(B_{\mu}+Q_\mu+\frac{i}{g}\partial_{\mu}+\delta_\mu^0\frac{1}{g}\omega \hat{L}_z)U^\dagger-B_\mu.
    \end{gathered}
    \label{GT2}
\end{equation}
and it is still necessary to perform gauge fixing. In the non-rotating case, according to the background-field method, using the Feynman gauge, the chosen gauge-fixing term satisfies
\begin{gather}
D_\mu(B)Q^\mu=\partial_\mu Q^\mu-ig[B_\mu,Q^\mu],\\
G[Q]^a=[D_\mu(B)Q^\mu]^a=(\delta^{ab}\partial_\mu-g\epsilon^{abc}B_\mu^{c})Q^{\mu,b},\\
    \mathcal{L}_\text{gf}(B,Q)=-\text{Tr}(D_\mu(B)Q^\mu)^2=-\frac{1}{2}G[Q]^aG[Q]^a,\\
    D_\mu(B^U)Q^{U,\mu}=UD_\mu(B)Q^\mu U^\dagger,\\
    \mathcal{L}_\text{gf}(B,Q)=\mathcal{L}_\text{gf}(B^U,Q^U).
\end{gather}
Here \(D_\mu=\partial_\mu+igB_\mu\). In the rotating case, the derivative operator \(\partial_\mu\rightarrow\partial_\mu+\delta_{\mu 0}R_0\), and because \(\delta_{\mu 0}R_0 Q^\mu=R_0 Q^0=-\delta_{\mu 0} \omega i\hat{L}_z Q^\mu\), \(R_0\) degenerates into \(-\omega i\hat{L}_z\). 
\begin{equation}
    D_\mu(B)Q^\mu=(\partial_\mu-\omega i\hat{L}_z\delta_{0\mu}) Q^\mu-ig[B_\mu,Q^\mu].\\
\end{equation}
\begin{widetext}
Using the background-field gauge transformation rules in Eqs.~(\ref{GT1}),
\begin{align}
     D_\mu(B^U)Q^{U,\mu}=&(\partial_\mu-\omega i\hat{L}_z\delta_{0\mu}) Q^{U,\mu}-ig[B_\mu^U,Q^{U,\mu}]\notag\\ 
     =&(\partial_\mu-\omega i\hat{L}_z\delta_{0\mu}) Q^{U,\mu}-ig[U(B_{\mu}+\frac{i}{g}\partial_{\mu}+\delta_\mu^0\frac{1}{g}\omega \hat{L}_z)U^\dagger ,Q^{U,\mu}]\notag\\
     =&(\partial_\mu-\omega i\hat{L}_z\delta_{0\mu}) Q^{U,\mu}-ig[UB_{\mu}U^\dagger ,Q^{U,\mu}]+[U(\partial_{\mu}-\omega i\hat{L}_z\delta_{0\mu})U^\dagger,Q^{U,\mu}]\notag\\ 
     =&U\left(\partial_\mu Q^\mu-ig[B_\mu,Q^\mu]-\omega i\hat{L}_z Q^{0}\right)U^\dagger\notag \\ 
     =&UD_\mu(B)Q^\mu U^\dagger.
\end{align}
\end{widetext}
This canonical angular momentum operator \(\hat{L}_z\), which is regarded as the term breaking gauge symmetry, is precisely an important term satisfying gauge symmetry.

The finite gauge transformation is \(U(x)\in SU(2)\). In the adjoint representation, it corresponds to a \(3\times 3\) orthogonal matrix \(R(U)\in SO(3)\), such that the transformation law of the adjoint field is
\begin{equation}
Q_\nu^{U,a}
=
R^{ab}(U)\,Q_\nu^b,
\end{equation}
As a gauge field, the transformation law of the background field \(B_\mu^a\) in the adjoint representation can be written as
\begin{equation}
B_\mu^{U,a}
=
R^{ab}B_\mu^b
-
\frac{1}{g}\,\left((\partial_\mu-\omega i\hat{L}_z\delta_{0\mu}) R^{ab}\right)(R^{T})^{bc},
\end{equation}
And from this one can obtain
\begin{equation}
    D_\mu(B^U)^{ab}=R(U)^{ac}D_\mu(B)^{cd}R(U)^{T\ db}.\label{GT3}
\end{equation}
For the corresponding functional determinant,
\begin{equation}
    \Delta_\text{F-P}[B,Q]=\det\left[ \frac{\delta G^a[Q^{\bar{U}}]}{\delta\omega^b}\Bigg|_{\bar{U}=I}\right],
\end{equation}
\begin{widetext}
under the infinitesimal transformation
\begin{gather}    U=\exp\left(i\omega(x)\right),\quad\omega(x)=\omega^a(x)\frac{\sigma^a}{2},\\
    \delta Q_\mu^{\bar{U}}=\frac{1}{g}\left( (\partial_\mu-\omega i\hat{L}_z\delta_{0\mu}) \omega -ig\left[A_\mu,\omega\right]\right)=\frac{1}{g}D_\mu(B+Q)\omega,\\
    \delta G^a[Q^{\bar{U}}]=[D_\mu(B)\delta Q^{\bar{U},\mu}]^a=\frac{1}{g}[D_\mu(B)D^\mu(B+Q)\omega]^a.
\end{gather}
Using Eq.~(\ref{GT3}), we obtain
\begin{equation}
    \Delta_\text{F-P}[B,Q]=\det[D_\mu(B)D^\mu(B+Q)]^{ab}=\det[D_\mu(B^U)D^\mu(B^U+Q^U)]^{ab}=\Delta_\text{F-P}[B^U,Q^U].
\end{equation}
In the one-loop calculation, we keep
\begin{equation}
    \Delta_\text{F-P}[B]=\det[D_\mu(B)D^\mu(B)]^{ab}=\det\left[-\left((\partial_\mu-\omega i\hat{L}_z\delta_{0\mu})+igB_\mu\right)\left((\partial^\mu-\omega i\hat{L}_z\delta_{0\mu})+igB^\mu\right)\right]^2.
\end{equation}
This term will cancel the contribution of the longitudinal polarization part in \(\mathcal{L}_\text{1-loop}\).
Thus we have proved that, in our article, the Lagrangian as well as the gauge-fixing functional determinant both satisfy background-field gauge invariance. Therefore,
\begin{align}
    Z^{(1)}[B]=\ln&\int \mathcal{D}[Q] \Delta_\text{F-P}[B]\exp{\left(S_\text{1-loop}[B,Q]+S_\text{gf}[B,Q]\right)}\notag\\ 
    =\ln&\int \mathcal{D}[Q] \Delta_\text{F-P}[B^U]\exp{\left(S_\text{1-loop}[B^U,Q^U]+S_\text{gf}[B^U,Q^U]\right)}\notag\\ 
    =\ln&\int \mathcal{D}[Q] \Delta_\text{F-P}[B^U]\exp{\left(S_\text{1-loop}[B^U,Q]+S_\text{gf}[B^U,Q]\right)}.
\end{align}
where the integration measures are equal, \(\mathcal{D}[Q^U]=\mathcal{D}[Q]\), and the whole calculation is completely gauge invariant with respect to the background field \(B\), namely 
\[
Z^{(1)}[B]=Z^{(1)}[B^U].
\]
\section{\MakeUppercase{Causality bound in the rotating frame}}
\label{B}
The inclusion of boundary conditions deforms the Landau wave functions due to finite-size effects:

\begin{equation}
    \tilde{\Phi}_\lambda^{(l)}(\rho,\phi)=\frac{1}{|l|!}\sqrt{\frac{gH}{2\pi}}\sqrt{\frac{\Gamma(\lambda+|l|+1)}{\Gamma(\lambda+1)}} F(-\lambda,|l|+1,X)\cdot X^{|l|/2}e^{-\frac{1}{2}X}e^{il\phi}.
\end{equation}
Here, $F(-\lambda,|l|+1,x)$ represents the confluent hypergeometric function, also known as Kummer's function of the first kind. It satisfies the boundary condition:
\begin{equation}
    F(-\lambda,|l|+1,X)=0,\quad X=\frac{1}{2}gH R^2.
\end{equation}
The quantum number $\lambda$ is thereby discretized into $\lambda_k^{|l|}$, denoting the $k$-th root of $F(-\lambda,|l|+1,X)=0$. The orthogonality condition is given by:
\begin{equation}
    \int_0^Xdx F(-\lambda_k^{|l|},|l|+1,x)F(-\lambda_{k'}^{|l|},|l|+1,x) x^{|l|}e^{-x}=N\delta_{kk'}.
\end{equation}
We define the normalization constant for the radial part as:
\begin{equation}
    N_k^{|l|}=\frac{1}{2\pi}\int_0^{2\pi}d\phi\int_0^R \rho d\rho |\tilde{\Phi}_{\lambda_k^{|l|}}^{(l)}(\rho,\phi)|^2 
    =\frac{1}{2\pi}\frac{1}{|l|!^2}\frac{\Gamma(\lambda+|l|+1)}{\Gamma(\lambda+1)}\int_0^Xdx F(-\lambda_k^{|l|},|l|+1,x)^2 x^{|l|}e^{-x}.
\end{equation}
To compute the spectral density $\sum_\lambda\braket{\lambda}{\lambda}$, we note that the quantum state of the system is described by $\ket{\lambda}=\ket{\lambda_k^{|l|},l,k_z} $. We have:
\begin{equation}
    \braket{\lambda_{k'}^{|l'|},l',k_z'}{\lambda_k^{|l|},l,k_z}=\int d^3\vec{x}\braket{\lambda_k^{|l|},l,k_z}{\vec{x}}\braket{\vec{x}}{\lambda_k^{|l|},l,k_z} =2\pi \delta_{l,l'}\cdot2\pi \delta(k_z-k_z')\cdot N_k^{|l|}.
\end{equation}
Using $\sum_\lambda\braket{\lambda'}{\lambda}=1$, we obtain the spectral density:
\begin{equation}
    \sum_{\lambda}=\frac{1}{2\pi}\sum_{l}\sum_k (N_k^{|l|})^{-1}\int_{-\infty}^{+\infty}\frac{dk_z}{2\pi}.
\end{equation}
The finite-temperature potential is given by:
\begin{align}
    V_T=&\sum_{s=\pm}\frac{T}{2\pi}\sum_{l}\sum_k \frac{|\tilde{\Phi}_{\lambda_k^{|l|}}^{(l)}(\rho,\phi)|^2}{N_k^{|l|}}\int_{-\infty}^{+\infty}\frac{dk_z}{2\pi} \ln \left(1-e^{-\beta\omega_s+i[\phi+(l-s)\tilde{\Omega}]}\right)+\ln \left(1-e^{-\beta\omega_s-i[\phi+(l-s)\tilde{\Omega}]}\right)_{\Omega=-i\omega}\notag\\
    =&\sum_{s=\pm}\frac{T}{2\pi}\sum_{l}\sum_k \frac{|\tilde{\Phi}_{\lambda_k^{|l|}}^{(l)}(\rho,\phi)|^2}{N_k^{|l|}}\int_{-\infty}^{+\infty}\frac{dk_z}{2\pi} \ln \left(1-e^{-\beta[\omega_s-(l-s)\omega]+i\phi}\right)+\ln \left(1-e^{-\beta[\omega_s+(l-s)\omega]-i\phi}\right),
\end{align}
\end{widetext}
with the corresponding eigen-spectrum:
\begin{equation}
    \omega_s=\sqrt{k_z^2+(2\lambda_k^{|l|}+2s+|l|-l+1)gH}.
\end{equation}
To study the thermodynamic potential at $r=0$, we utilize the property:
\begin{equation}
    \lim_{\vec{r}\rightarrow0 }|\tilde{\Phi}_{\lambda_k^{|l|}}^{(l)}(\rho,\phi)|^2=\frac{gH}{2\pi}\delta_{l,0},
\end{equation}
from which we can derive the thermodynamic potential in Eq.~(\ref{thermo}).
\section{\MakeUppercase{the limit of vanishing chromomagnetic field}}
\label{C}
We first examine the limit of vanishing chromomagnetic field, $H\rightarrow 0$, under an imaginary angular velocity. 
\begin{equation}
    \Delta k_{\perp}^2=2gH:\  \frac{gH}{2\pi}\sum_{m=0}^{+\infty}=\frac{gH}{2\pi}\sum_{m=0}^{+\infty}\frac{dk_{\perp}^2}{\Delta k_{\perp}^2}=\int_{0}^\infty\frac{dk_\perp^2}{4\pi}.
\end{equation}
The summation over Landau levels transforms into an integral. For the tachyonic mode, we have:
\begin{widetext}
\begin{equation}
    u(-,0,\vec{0})=\lim_{H\rightarrow 0}\frac{gH}{\pi\beta}\int_{-\infty}^{+\infty}\frac{dk}{2\pi} \operatorname{Re}\ln \left(1-e^{-\beta\sqrt{k^2-gH}+i(\phi+\tilde{\Omega})}\right) =\lim_{H\rightarrow 0}\frac{2gH}{\pi\beta}\int_{0}^{\sqrt{gH}}\frac{dk}{2\pi}\operatorname{Re}\ln \left(1-e^{-\beta\sqrt{k^2-gH}+i(\phi+\tilde{\Omega})}\right).
\end{equation}
\end{widetext}
Utilizing the mean value theorem,
\begin{gather}
    \int_a^bdx f(x)=f(\epsilon)(b-a), \qquad \epsilon\in[a,b], \\
    \lim_{x\rightarrow 0^+} x^3\ln(1-e^{-ix})=0,
\end{gather}
we find that $u(-,0,\vec{0})=0$. Similarly, $u(-,1,\vec{0})=0$. Consequently, the finite-temperature thermodynamic potential reduces to:
\begin{align}
    V_T(\vec{0})&=T\sum_{s=\pm}\int \frac{dk_\perp^2 dk_z}{8\pi}2\operatorname{Re}\ln \left(1-e^{-\beta|\vec{k}|+i(\phi-s\tilde{\Omega})}\right)\notag\\
    &=2T\int\frac{4\pi k^2dk}{(2\pi)^3}-\sum_{n=1}^\infty\frac{1}{n}e^{-n\beta k}\sum_{s=\pm}\cos{n(\phi-s\tilde{\Omega})}\notag\\
    &=-\frac{2T^4}{\pi^2}\sum_{s=\pm}\sum_{n=1}^\infty\frac{1}{n^4}\cos{n(\phi-s\tilde{\Omega})}\notag\\
    &=\frac{2\pi^2T^4}{3}\sum_{s=\pm}B_4\left[\frac{1}{2\pi}(\phi-s\tilde{\Omega})\right]_{\mod 1}.
\end{align}

Next, we examine the limit $H\rightarrow 0$ under a real angular velocity. In the limit $X \rightarrow 0$, we have \cite{Chen:2017xrj}:
\begin{equation}
    e^{-X/2}F(-\lambda,1,X)\rightarrow J_0(\sqrt{4\lambda X}).
\end{equation}
The zeros satisfy:
\begin{equation}
     \quad \lambda_k^0\rightarrow \frac{{\epsilon_k^0}^2}{4X}, \qquad J_0(\epsilon_k^0)=0.
\end{equation}
The normalization constant becomes:
\begin{align}
    N_k^0=&\frac{1}{2\pi}\int_0^Xdx F(-\lambda_k^0,1,x)^2e^{-x}\notag\\
    =&\frac{1}{2\pi}\int_0^1 Xd\hat{r}^2 F(-\lambda_k^0,1,X\hat{r}^2)^2e^{-X\hat{r}^2}\notag\\
    \rightarrow&\frac{X}{\pi}\int_0^1\hat{r}d\hat{r}J_0(\epsilon_k^0\hat{r})^2=\frac{X}{2\pi}J_1^2(\epsilon_k^0).
\end{align}
The energy spectrum simplifies to:
\begin{equation}
    \omega_s^0=\sqrt{k_z^2+(2\lambda_k^0+2s+1)gH}\rightarrow\sqrt{k_z^2+(\epsilon_k^0/R)^2},
\end{equation}
which no longer depends on spin and is denoted as $\omega^0$. Substituting into Eq. (\ref{thermo}), we obtain the thermodynamic potential in the limit of a vanishing chromomagnetic field:
\begin{equation}
    V(\vec{0})=\frac{T}{\pi^2R^2}\sum_{s,k}\frac{1}{J_1^2(\epsilon_k^0)}\int_{-\infty}^{+\infty}dk_z\ln\big|1-e^{-\beta(\omega^0+s\omega)+i\phi}\big|.
\end{equation}

\bibliography{ref}
\end{document}